\documentclass[traditabstract]{aa}
\usepackage{graphicx,textcomp,amsmath,txfonts}
\usepackage{natbib}
\bibpunct{(}{)}{;}{a}{}{,}

\begin{document}

\sloppy

%-----------------------------------------------------------------------
% Authors' Macros
%-----------------------------------------------------------------------

\renewcommand*{\phi}{\varphi}
\renewcommand*{\epsilon}{\varepsilon}
\newcommand{\total}{\operatorname{d}\!}
\renewcommand{\arraystretch}{1.05}
\tabcolsep 3pt

\newcommand{\mum}{\,\hbox{\textmu{}m}}
\newcommand{\mm}{\,\hbox{mm}}
\newcommand{\cm}{\,\hbox{cm}}
\newcommand{\m}{\,\hbox{m}}
\newcommand{\km}{\,\hbox{km}}
\newcommand{\AU}{\,\hbox{AU}}
\newcommand{\pc}{\,\hbox{pc}}
\newcommand{\Kelvin}{\,\hbox{K}}
\newcommand{\second}{\,\hbox{s}}
\newcommand{\yr}{\,\hbox{yr}}
\newcommand{\Myr}{\,\hbox{Myr}}
\newcommand{\Gyr}{\,\hbox{Gyr}}
\newcommand{\rad}{\,\hbox{rad}}
\newcommand{\mJy}{\,\hbox{mJy}}
\newcommand{\magn}{\,\hbox{mag}}
\newcommand{\gperccm}{\,\hbox{gcm}^{-3}}
\newcommand{\ergperg}{\,\hbox{ergg}^{-1}}
\newcommand{\gpers}{\,\hbox{gs}^{-1}}
%------------------------------------------------------------------

\title{The Edgeworth-Kuiper debris disk}

\bigskip
\author{Christian Vitense\inst{1}
        \and
        Alexander V. Krivov\inst{1}
        \and
        Torsten L\"ohne\inst{1}
       }
\offprints{A.V.~Krivov, \email{krivov@astro.uni-jena.de}}
\institute{Astrophysikalisches Institut, Friedrich-Schiller-Universit\"at Jena,
           Schillerg\"a{\ss}chen~ 2--3, 07745 Jena, Germany
          }
\date{Received February 5, 2010; accepted {\em date}}

\abstract
{
The Edgeworth-Kuiper belt (EKB) and its presumed dusty debris is
a natural reference for extrsolar debris disks.
We re-analyze the current database of known transneptunian objects (TNOs)
and employ a new algorithm to eliminate the inclination and the distance selection
effects in the known TNO populations to derive expected
parameters of the ``true'' EKB.
Its estimated mass is $M_\text{EKB} = 0.12 M_\oplus$,
which is by a factor of $\sim 15$ larger than the mass
of the EKB objects detected so far.
About a half of the total EKB mass is in classical and resonant
objects and another half is in scattered ones.
Treating the debiased populations
of EKB objects as dust parent bodies, we then ``generate'' their dust disk
with our collisional code.
Apart from accurate handling of
destructive and cratering collisions and direct radiation pressure, we
include the Poynting-Robertson (P-R) drag.
The latter is known to be
unimportant for debris disks around other stars detected so far, but cannot
be ignored for the EKB dust disk because of its much lower optical depth.
We find the radial profile of the normal optical depth to peak
at the inner edge of the classical belt, $\approx 40\AU$.
Outside the classical EKB, it approximately follows
$\tau \propto r^{-2}$
which is roughly intermediate between the slope
predicted analytically for collision-dominated ($r^{-1.5}$)
and transport-dominated ($r^{-2.5}$) disks.
The size distribution of dust is less affected by the P-R effect.
The cross section-dominating grain size still lies just above
the blowout size ($\sim 1 \dots 2\mum$), as it would if the P-R effect was ignored.
However, if the EKB were by one order of magnitude less massive, its dust disk would have
distinctly different properties. The optical depth profile would fall off
as $\tau \propto r^{-3}$, and
the cross section-dominating grain size 
would shift from $\sim 1\dots 2\mum$ to $\sim 100\mum$.
These properties are seen if dust is assumed to be generated only by known TNOs
without applying the debiasing algorithm.
An upper limit of the in-plane optical depth of the EKB dust set by our model
is $\tau = 2 \times 10^{-5}$ outside $30\AU$.
If the solar system were observed from outside, the thermal emission flux from the EKB dust
would be about two orders of magnitude lower than for
solar-type stars with the brightest known infrared excesses observed from the same distance.
Herschel and other new-generation facilities should reveal extrasolar debris disks
nearly as tenuous as the EKB disk.
We estimate that the Herschel/PACS instrument should be able to detect disks at a 
$\sim 1 \dots 2M_\text{EKB}$ level.


}

\authorrunning{Vitense et al.}
\titlerunning{The Edgeworth-Kuiper debris disk}

\maketitle

%------------------------------------------------------------------
\section{Introduction}

Debris disks, now known to be ubiquitous around main-sequence stars,
are the natural aftermath of the evolution of dense protoplanetary disks
that may or may not result in formation of planets
\citep[see, e.g.,][for recent reviews]{Wyatt-2008,Krivov-2010}.
They are composed of left-over planetesimals and smaller debris produced
in mutual collisions, and it is the tiniest, dust-sized collisional fragments
that are evident in observations through thermal radiation
and scattered stellar light.

Like planetary systems of other stars, our solar system contains planetesimals
that have survived planetary formation.
The main asteroid belt between two groups of planets, terrestrial and giant ones,
comprises planetesimals that failed to grow to planets because of the strong perturbations
by the nearby Jupiter \citep[e.g.,][]{Safronov-1969,Wetherill-1980}.
The Edgeworth-Kuiper Belt (EKB) exterior
to the Neptune orbit is built up by planetesimals that did not form planets
because the density of the outer solar nebula was too low
\citep[e.g.,][]{Safronov-1969,Lissauer-1987,Kenyon-Bromley-2008}.
Both the asteroid belt and the Kuiper belt are heavily structured dynamically, 
predominantly by Jupiter and Neptune respectively. They
include non-resonant and resonant families,
as well as various objects in transient orbits ranging from detached and scattered
Kuiper-belt objects through Centaurs to Sun-grazers.
Short-period comets, another tangible population of small bodies in the inner
solar system, must be genetically related to the Kuiper belt that acts as their
reservoir \citep{Quinn-et-al-1990}.
Asteroids and short-period comets together are sources of interplanetary
dust, observed in the planetary region,
although their relative contribution to the dust production remain uncertain 
\citep{Gruen-et-al-2001}.
And this complex system structure was likely quite different in the past.
It is argued that the giant planets and the Kuiper belt have originally formed in
a more compact configuration (the ``Nice model'',
\citeauthor{Gomes-et-al-2005} \citeyear{Gomes-et-al-2005}), and that
it went through a short-lasting period of dynamical instability, likely
explaining the geologically recorded event of the Late Heavy Bombardment (LHB).

As the amount of material and spatial dimensions of the EKB surpass by far
those of the asteroid belt and the population of short-period comets, it is the EKB
and its presumed collisional debris that should be referred to as the debris disk
of the solar system. Ironically, the observational status of the solar system's debris disk
is the opposite of that of the debris disks around other stars.
In the latter case, as mentioned above, it is dust that can be observed.
In the former case, we can observe the planetesimals, but there is no certain detection
of their dust so far \citep{Gurnett-et-al-1997,Landgraf-et-al-2002}.

An obvious difference between the debris disks detected so far around other stars
and our solar system's debris disk is the total mass (and thus, also the amount
of dust).
\citet{Mueller-et-al-2009} for example infer several Earth 
masses as the total mass of the Vega debris disk, whereas the Kuiper belt mass is reported to be
below one-tenth of the Earth mass \citep{Bernstein-et-al-2004,Fuentes-Holman-2008}.
As a result, were the solar system observed from afar, its debris dust would be
far below the detection limits. However, a number of debris disks around
Sun-like stars that are coeval with or even older than the Sun have been detected.
\citet{Booth-et-al-2009} analyze ``dusty consequences'' of a major depletion of the 
planetesimal populations in the solar system at the LHB phase. They point out
that the pre-LHB debris disk of the Sun would be among the brightest debris
disks around solar-type stars currently observed. Future, more sensitive
observations (for instance, with Herschel Space Observatory) should detect lower-mass
disks, bridging
the gap between dusty debris disks
around other stars and tenous debris disks in the present-day solar system.

Given the low mass of the dust parent bodies in the EKB, our debris disk
is thought to fall into the category of the so-called transport-dominated disks
\citep{Krivov-et-al-2000},
where radial transport timescales for dust (due to the Poynting-Robertson effect,
henceforth P-R, e.g. \citeauthor{Burns-et-al-1979} \citeyear{Burns-et-al-1979})
 are shorter than their collisional lifetime.
This is opposite to extrasolar debris disks that are collision-dominated
\citep{Wyatt-2005}.
Whereas the latter have been extensively modeled both analytically and 
numerically
\citep{Thebault-et-al-2003,Krivov-et-al-2006,Strubbe-Chiang-2006,Thebault-Augereau-2007,Wyatt-et-al-2007,Loehne-et-al-2008,Mueller-et-al-2009},
more effort has to be invested into modeling of transport-dominated disks.
The more so, as new-generation facilities like Herschel Space Observatory will soon open a phase
when low-density, transport-dominated disks can be studied observationally.

The main goal of this paper is to develop a more realistic model of such
a tenuous disk, exemplified by the EKB dust disk,
than was done before \citep[e.g.][]{Stern-1995}.
We take an advantage that~--- unlike with other debris disks and unlike at the time
when the first collisional models of the EKB dust were devised~---
more than a thousand EKB objects, acting as dust parent bodies of the solar system's debris 
disk, are now known.
This task is accomplished in two steps:
\begin{itemize}
\item[I.]
First, the currently known objects in the EKB are analyzed.
In Sect. 2, we work out an algorithm to correct their distributions
for observational selection effects and try to reconstruct the properties
of the expected ``true'' EKB.
\item[II.]
Second, we treat the objects of the ``true'' EKB as dust parent bodies.
In Sect. 3, we make simulations of dust production and evolution with a statistical code,
fully taking into account collisions and the P-R effect, and present
the expected radial and size distribution of the presumed EKB dust.
In Sect.~4, we model the spectral energy distribution (SED) of the simulated
EKB dust disk and compare it to the SEDs of other debris disks.
We also compare these results with the detection limits of the Herschel/PACS instrument.
\end{itemize}
Our results are summarized in Sect.~5 and discussed in Sect.~6.

%------------------------------------------------------------------

\section{Planetesimals in the Kuiper belt}

\subsection{Observations and their biases}

The EKB was predicted more than fifty years ago by Edgeworth and Kuiper and it took fourty
years until the first object, QB~1, was discovered \citep{Jewitt-et-al-1992}.
More than 1260 transneptunian objects (TNOs) orbiting the Sun beyond the orbit 
of Neptune have been discovered.
Table \ref{tab:observations} lists most of the surveys published so far, in which new TNOs
have been discovered, and key parameters of these surveys.
One parameter is the area $\Omega$ on the sky searched for TNOs.
Another one is the limiting magnitude $m_{50}$
that corresponds to the detection probability of $50\%$.
As the detection probability drops rapidly from 100\% to zero when the apparent 
magnitude $m$ ``crosses'' $m_{50}$, we simply assume that an object
will be detected with certainty if $m < m_{50}$ and missed otherwise.
Finally, the maximum ecliptic latitude $\epsilon$ covered by each 
survey is listed.
Where it was not given explicitly in the original papers, we estimated it
to be $\sqrt{\Omega}/2$, assuming
that the surveyed area was centered on the ecliptic.
Table \ref{tab:observations} shows that all campaigns
can be roughly divided into two groups:
deeper ones with a small sky area covered (``pencil-beam'' surveys)
and shallower ones with a larger area, but a smaller limiting magnitude.

\begin{table}[bt!]
\centering
\begin{tabular}{|c|c|c|c|c|c|}
\hline
$\Omega$ [deg$^2$] &	N 	& $m_{50}$	&	$\epsilon$ [${}^{\circ}$]&$\alpha$ [${}^{\circ}$]	& Reference\\
\hline\hline
0.7	&	2	&	23.5	&	$0.42$	&$0.84$&        \cite{Irwin-et-al-1995}$^\star$	\\
1.2	&	7	&	24.85	&	$0.55$	&$1.1$&		\cite{Jewitt-et-al-1995}	\\
3.9	&	12	&	24.2	&	$0.99$	&$1.97$&	\cite{Jewitt-et-al-1996}$^\star$	\\
4.4	&	3	&	23.2	&	$1.05$	&$2.1$&		\cite{Jewitt-et-al-1996}$^\star$	\\
0.35	&	1	&	24.6	&	$0.30$	&$0.59$&	\cite{Gladman-et-al-1998}	\\
0.049	&	4	&	25.6	&	$0.11$	&$0.22$&	\cite{Gladman-et-al-1998}$^\star$	\\
0.075	&	0	&	25	&	$0.14$	&$0.27$&	\cite{Gladman-et-al-1998}	\\
51.5	&	13	&	23.4	&	$3.6$	&$7.2$&		\cite{Jewitt-et-al-1998}	\\
0.01	&	2	&	27.94	&	$0.05$	&$0.1$&		\cite{Chiang-Brown-1999}	\\
20.2	&	3	&	23.6	&	$2.25$	&$4.5$&		\cite{Trujillo-et-al-2000}	\\
1.5	&	24	&	24.9$\dots$ 25.9&$0.61$	& $1.22$&	\cite{Allen-et-al-2001}$^\star$	\\
0.012	&	0	&	26.7	&	$0.06$	&$0.11$&	\cite{Gladman-et-al-2001}$^\star$	\\
0.31	&	17	&	25.93	&	$0.28$	&$0.56$&	\cite{Gladman-et-al-2001}$^\star$	\\
73	&	86	&	24.0	&	$4.25$	&$8.5$&		\cite{Trujillo-et-al-2001a}$^\star$	\\
164	&	3	&	21.1	&	$6.40$	&$12.8$&	\cite{Trujillo-et-al-2001b}$^\star$	\\
5108	&	19	&	20.7 	&	$10$	&$255.4$&	\cite{Trujillo-Brown-2003}$^\star$	\\
0.02	&	3	&	28.7	&	$0.07$	&$0.14$&	\cite{Bernstein-et-al-2004}$^\star$	\\
550	&	183	&	22.5	&	$5$	&$55$&		\cite{Elliot-et-al-2005}$^\star$	\\
8000	&	1 big	&	20$\dots$21&	$10$	&$400$&	 	\cite{Larsen-et-al-2007}$^\star$	\\
3.0	&	70	&	26.4	&	$0.87$	&$1.73$&	\cite{Fraser-et-al-2008}	\\
2.8	&	82	&	25.7	&	$1.67$	&$1.67$&	\cite{Fuentes-Holman-2008}	\\
0.255	&	20	&	26.76	&	$0.25$	&$0.5$& 	\cite{Fuentes-et-al-2009}	\\
0.33	&	36	&	26.8	&	$0.29$	&$0.57$&	\cite{Fraser-Kavelaars-2009}	\\
\hline
\end{tabular}
\vspace*{3mm}
\caption{A list of campaigns where TNOs were found.
The sky area covered ($\Omega$),
the number of the objects discovered ($N$),
the limiting magnitude ($m_{50}$),
an estimated half-opening angle $\epsilon$,
and the ecliptic longitude coverage $\alpha$
are given.
Papers that provide enough data for objects discovered in 
that survey to identify them in the MPC database
are marked with an asterisk.
}
\label{tab:observations}
\end{table}

The orbits of TNOs are commonly characterized by
six orbital elements:
semi-major axis $a$
(or perihelion distance $q$),
eccentricity $e$,
inclination $i$,
argument of pericenter $\omega$,
longitude of the ascending node $\Omega$,
and mean anomaly $M$.
In addition, each object itself is characterized by the absolute magnitude $H$, 
which is defined as the apparent magnitude the object would have if it was 
$1\AU$ away from the Sun and the Earth, and depends on the object radius and 
albedo.
We take the orbital elements and the absolute magnitudes of all known objects
from the Minor Planet Center (MPC) 
database\footnote{\textit{http://www.cfa.harvard.edu/iau/lists/TNOs.html} [Last accessed 12 October 2009]}
rather than from discovery papers listed in Tab.~\ref{tab:observations},
because the MPC data include follow-up observations and thus provide
a better precision.

\begin{figure*}
  \begin{center}
  \includegraphics[width=0.90\textwidth]{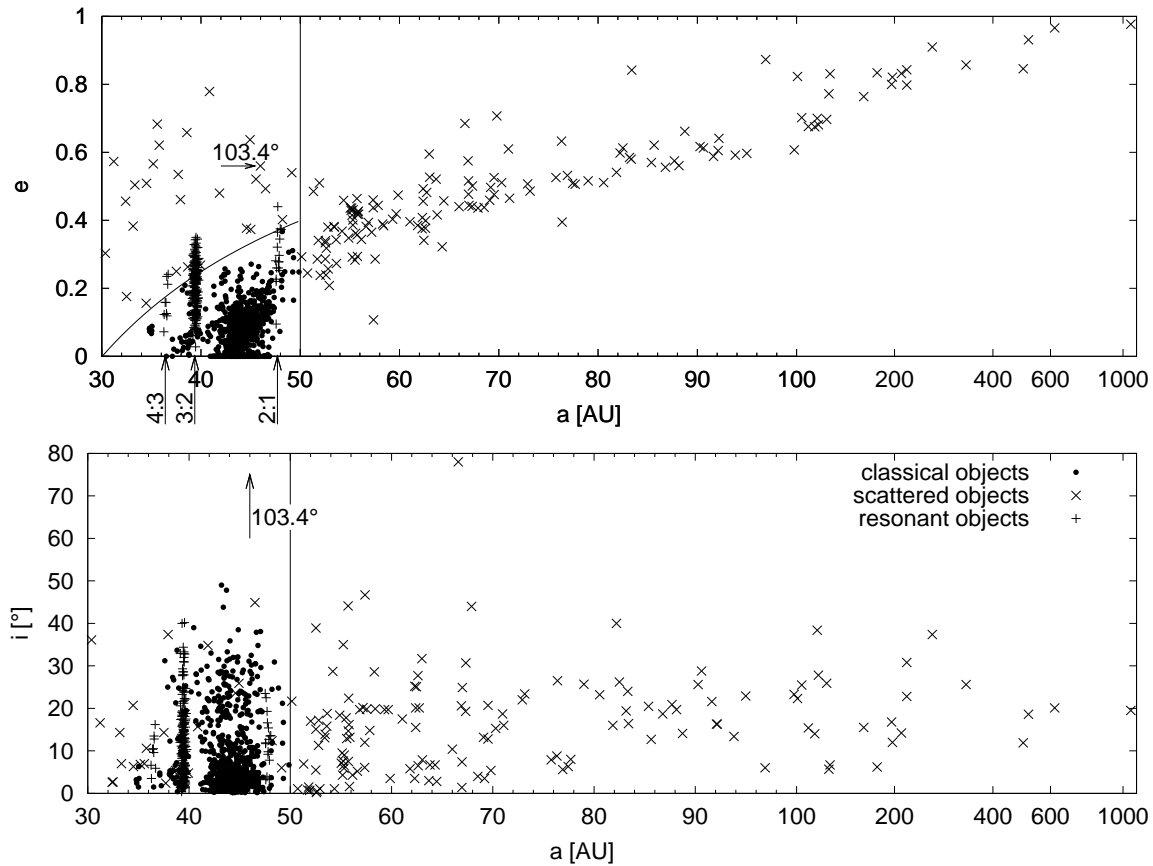}\\
  \end{center}
  \caption{Known TNOs in the $a$--$e$ plane (top) and $a$--$i$ plane (bottom).
  Different groups are shown with different symbols:
  $865$ classical objects with dots,
  $235$ resonant TNOs with pluses,
  and the remaining $160$ scattered objects with crosses.
  Solid lines separate classical and scattered objects in our classification.
  One object with $i=103.4^\circ$
  is outside the lowest panel, but is marked with arrows in both panels.
  Notice that the linear scale turns into a logarithmic scale at $100\AU.$
  }
  \label{fig:a-e-i-dist}
\end{figure*}

The planet formation theory implies that the TNO orbits strongly concentrate
towards the ecliptic plane.
Accordingly, in order to increase the detection probability,
the majority of the observations were made near the ecliptic, and only a few
surveys covered high ecliptic latitudes.
Our sample, given in Tab.~\ref{tab:observations},
contains surveys with  $\epsilon$ up to $10^\circ$
\citep[e.g.][]{Trujillo-et-al-2001a,Trujillo-Brown-2003,Elliot-et-al-2005,Petit-et-al-2006,Larsen-et-al-2007}.
However, TNOs with higher orbital inclinations exist as well.
Since \cite{Brown-2001} it is known that the inclination distribution of TNOs
has a second maximum at higher inclinations.
Several objects with very high inclinations, including one with $i=103.4^\circ$,
were detected.
Clearly, the fact that observations are done near the ecliptic plane decreases
the probability to detect such objects, because it is only possible twice per 
orbital period, close to the nodes.
Thus there is an obvious selection effect in favor of TNOs in low-inclination orbits
that needs to be taken into account.
Equally obvious is another selection effect, which is that objects are predominantly
discovered at smaller heliocentric distances. This reduces the probability to
discover TNOs with large semi-major axes and high eccentricities, because these are 
too faint all the time except for the short period of time when they are near perihelion.

\subsection{Classification of TNOs}

Many classifications of TNOs into ``classical'', 
``resonant'', ``excited'', ``scattered'', ``detached'' etc. groups have been proposed,
based on the orbital elements and taking into account dynamical arguments
\citep[e.g.,][among others]{Jewitt-et-al-1998,Chiang-Brown-1999,Luu-Jewitt-2002,Delsanti-Jewitt-2006,Jewitt-et-al-2009}.
Classifications by different authors are similar, but not identical.
In this paper we use the following working classification:
\begin{enumerate}
  \item Resonant objects (RES): objects in a mean-motion commensurability with 
        Neptune, where we only consider the three most prominent first-order 
        resonances 4:3, 3:2 and 2:1. To identify the objects as resonant,
        we use the resonance ``widths'' from \citet{MurrayDermott}. For example,
        the width of the $3:2$-resonance at $e=0.1$ is $\Delta a = 0.012\AU$.
        The width increases with increasing eccentricity and with decreasing 
        distance to Neptune.
  \item Classical Kuiper Belt (CKB) objects: objects with $a < 50\AU$,
        which are neither resonant nor
        Neptune-crossers ($q > a_\mathrm{Neptune}$).
  \item Scattered disk objects (SDO): objects with $a > 50\AU$, as well as Neptune-crossers
        ($a>a_\mathrm{Neptune}$ and $q<a_\mathrm{Neptune}$).
\end{enumerate}
Figure~\ref{fig:a-e-i-dist} depicts all known TNOs, using different symbols
for each of the three groups.
This classification is intentionally made simpler than many others in common use,
in order to facilitate the analysis below.
For instance, in our classification, the known detached objects
(those with perihelia well outside Neptune's orbit, $q>40\AU$)
fall into the ``scattered'' category.
On any account, the parameters
of the entire EKB and its dust that we will derive will not depend on the way
in which the TNOs are classified into various groups.
On the other hand, this classification roughly reflects different formation 
history of different populations in the EKB, as well as different modes of their
gravitational interaction with Neptune at present.

\subsection{Debiasing procedure}

Because of the obvious selection effects of inclined and faint objects,
statistical models were developed to estimate a true distribution of orbital
elements and numbers of the TNOs. \cite{Brown-2001} calculated an inclination
distribution. He assumed circular orbits and derived a relation between the
inclination and the fraction of an object's orbit that it spends at
low ecliptic latitudes. \citet{Donnison-2006} calculated the
magnitude distribution for the classical, resonant, and scattered objects for absolute 
magnitudes $H<7$,
using maximum likelihood estimations, and showed that the samples
were statistically different. 

Here we propose another debiasing method to estimate the ``true''
distribution of the TNOs, based on the obervational surveys listed in Table
\ref{tab:observations}. We start with calculating the probability to find an
object with the given parameters $\{a,e,i,\omega,\Omega,H\}$ for each given
survey.
To this end, we estimate the time fraction of the object's orbit that lies within
the maximum ecliptic latitude $\epsilon$ covered by the survey, as
well as the fraction of the orbit which is observable at the given limiting
magnitude $m_{50}$, and find the intersection of the two orbital arcs. Once
the probability to detect the object in each of the surveys has been
calculated, we compute the probability that it would be detected at least
in one of the surveys made so far.
Finally, we augment the number of objects with that same orbital elements
as the object considered to a $100\%$ probability, e.g., an
object with $20\%$ probability is counted five times.

We now explain this procedure in more detail. The first effect is the {\it
``inclination bias''}. In calculating the orbital arc that lies in the
observable latitudinal zone, we make the assumption that we observe from the sun.
The observable area on the sky is thus a belt $|b| \le \epsilon$, where $b$ is
the heliocentric ecliptic latitude.
The orbit crosses the boundary of the observed belt, $|b| = \epsilon$,
at four points. At these intersection points, the true anomaly $\phi$ takes
the values
\begin{equation}\label{eq:phi}
	\phi_j = \pm \arccos \left( \pm \sqrt{1-\frac{\sin^2\epsilon}{\sin^2 i}}\right) - \omega.
\end{equation}
Due to our approximation that we observe from the Sun, 
the longitude of the ascending node does not appear in this formula.

The second effect is the {\it ``distance bias''} (or {\it ``eccentricity
bias''}). The maximum distance at which an object is detectable is given by
\citep{Irwin-et-al-1995}
\begin{equation}\label{eq:rmax}
	r_\mathrm{max} = 10^{0.1(m_{50} - H)}.
\end{equation}

\begin{figure}
  \begin{center}
  \includegraphics[width=0.16\textwidth]{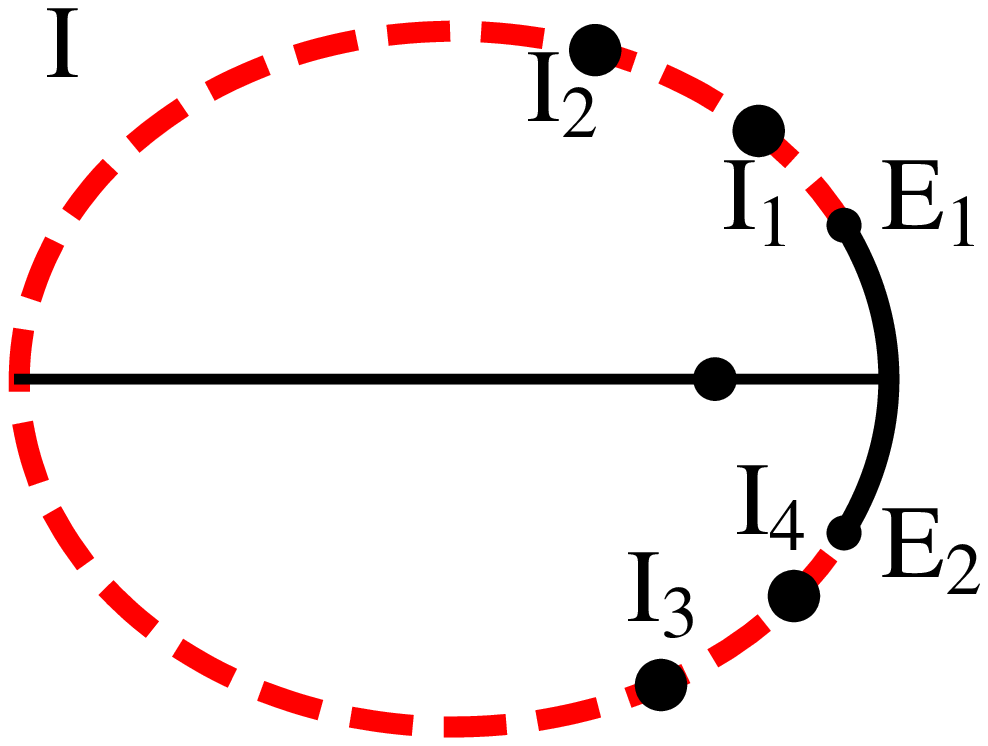}
  \includegraphics[width=0.16\textwidth]{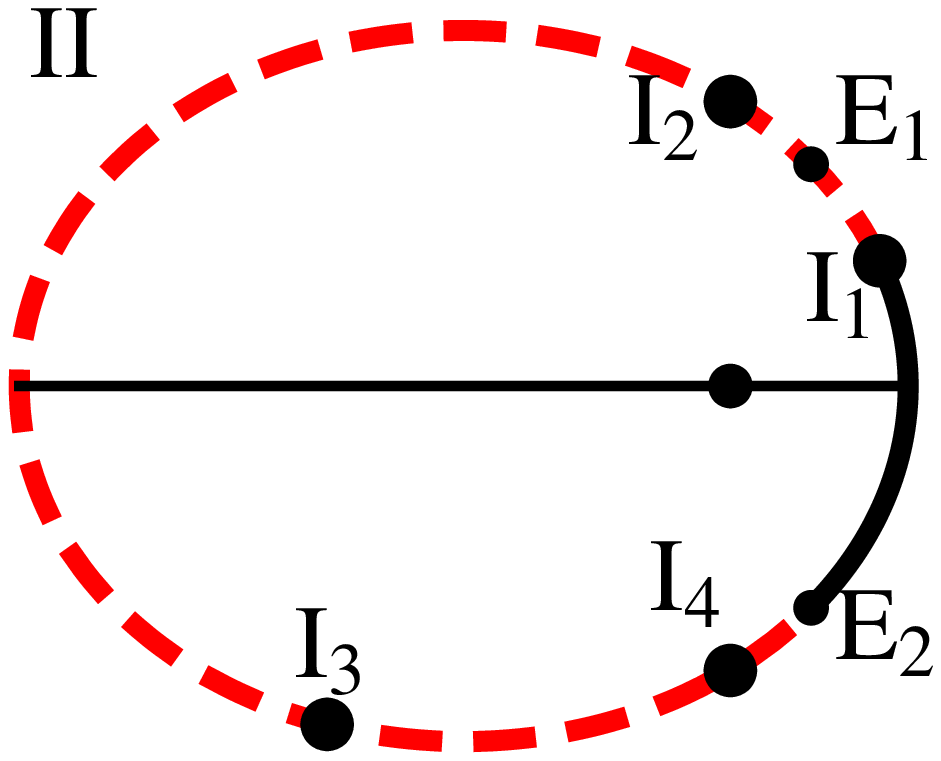}
  \includegraphics[width=0.16\textwidth]{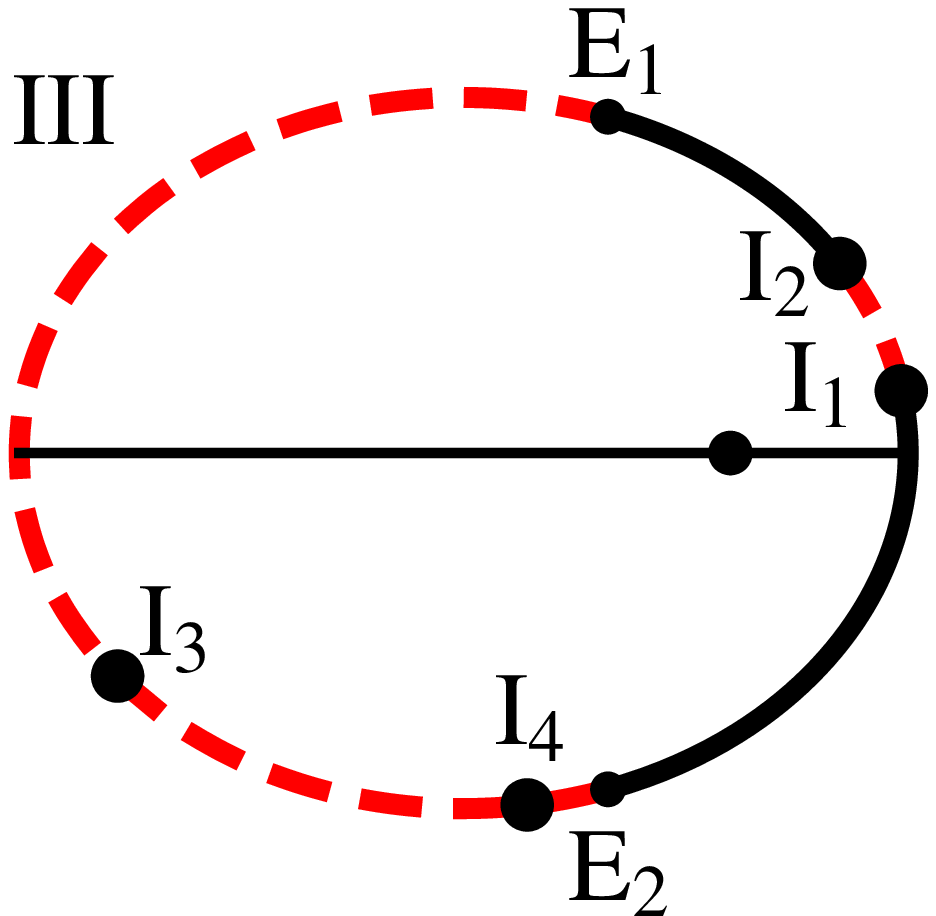}\\
  \includegraphics[width=0.16\textwidth]{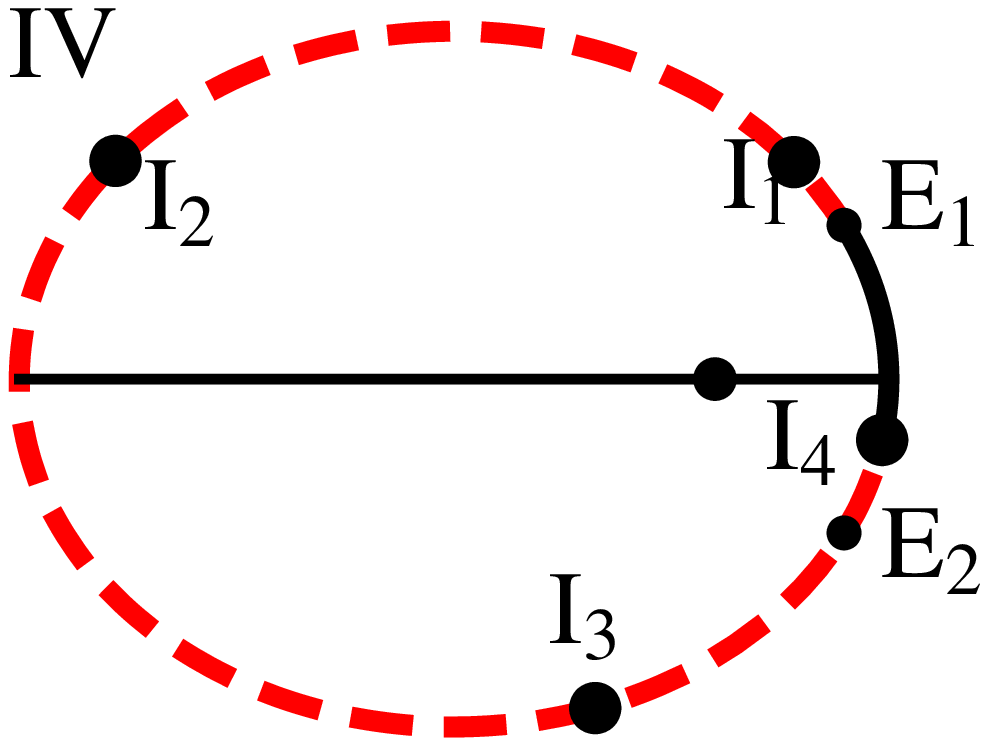}
  \includegraphics[width=0.16\textwidth]{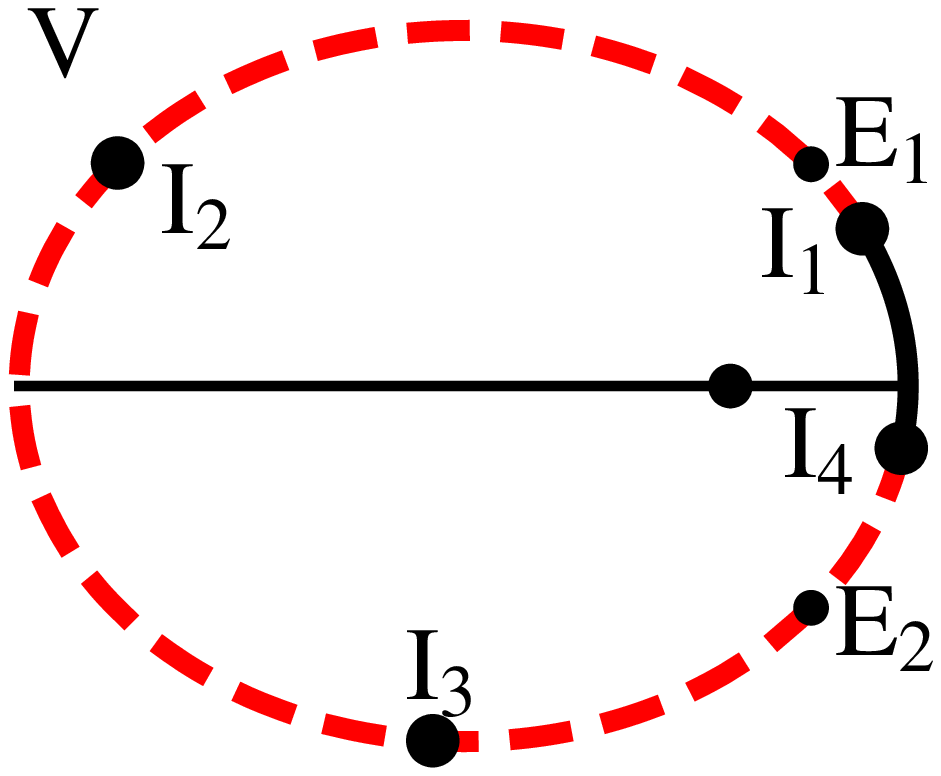}
  \includegraphics[width=0.16\textwidth]{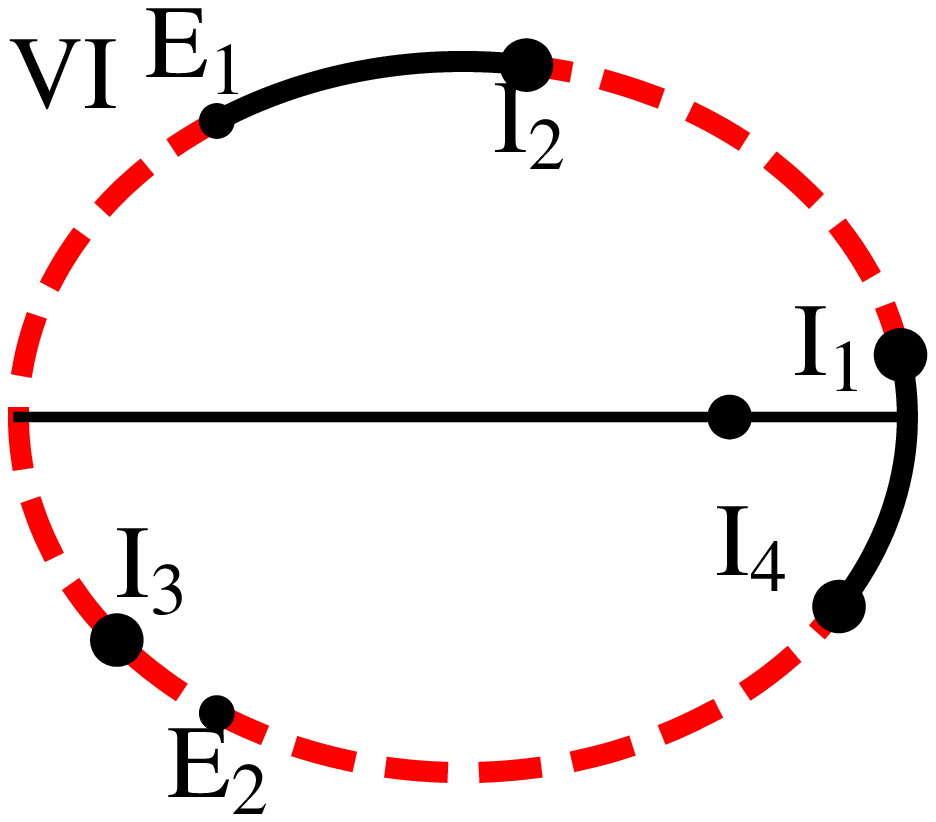}\\
  \includegraphics[width=0.16\textwidth]{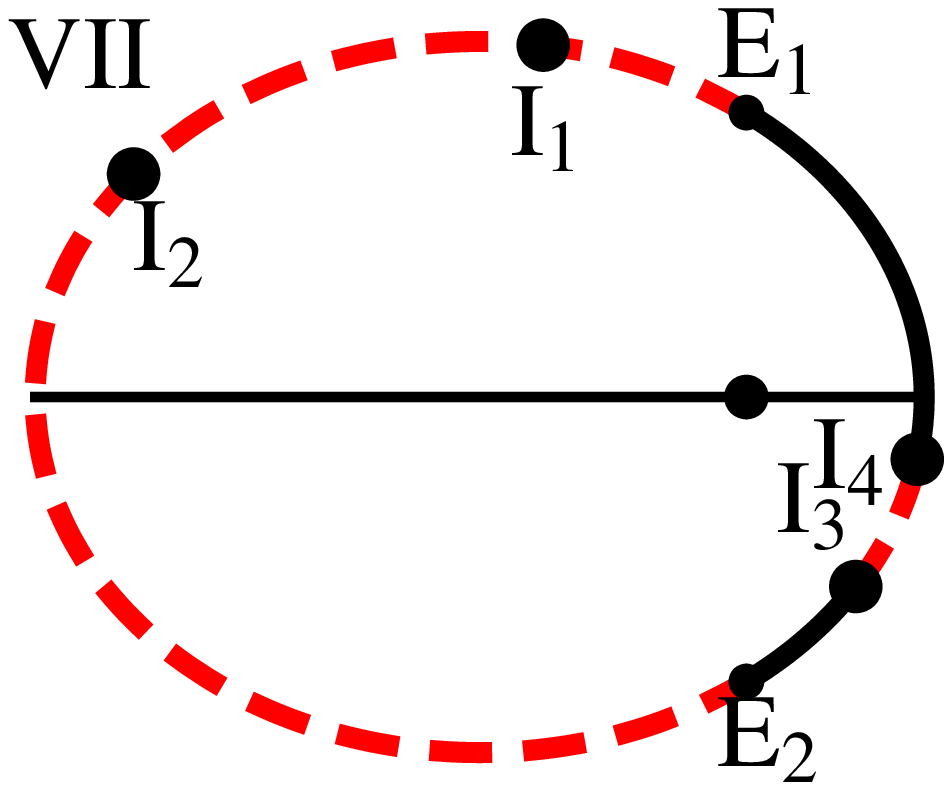}
  \includegraphics[width=0.16\textwidth]{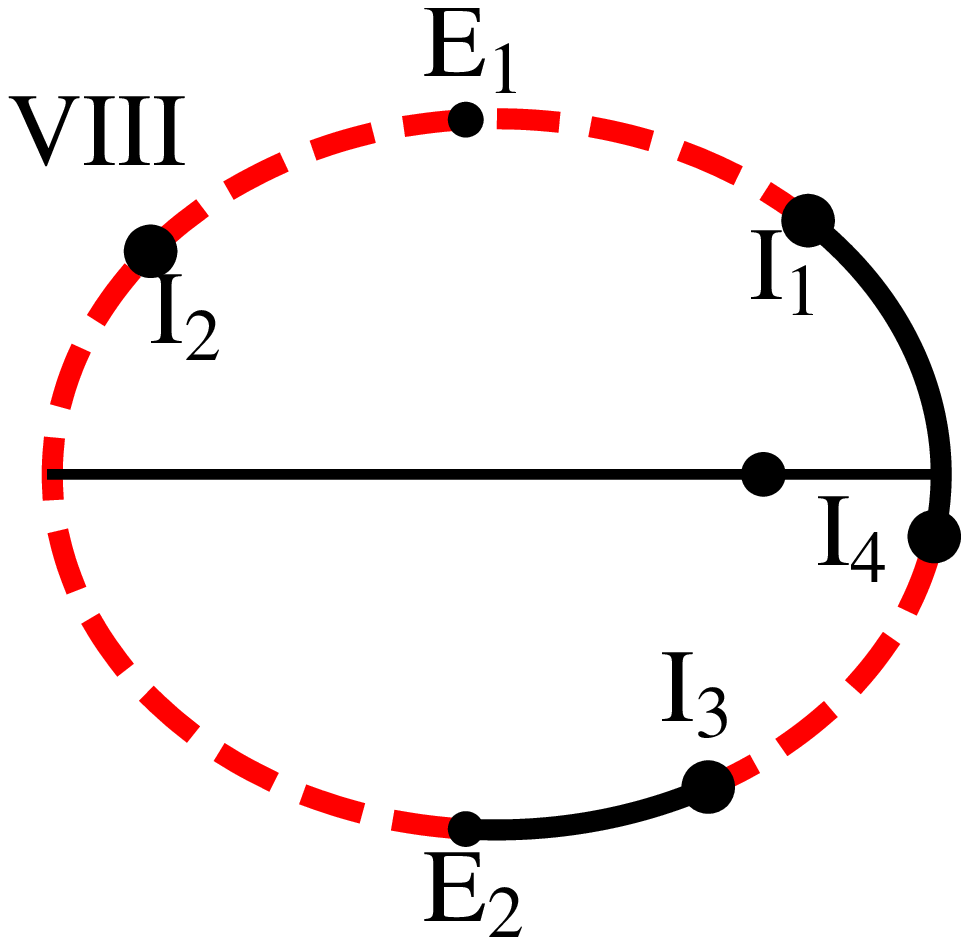}
  \includegraphics[width=0.16\textwidth]{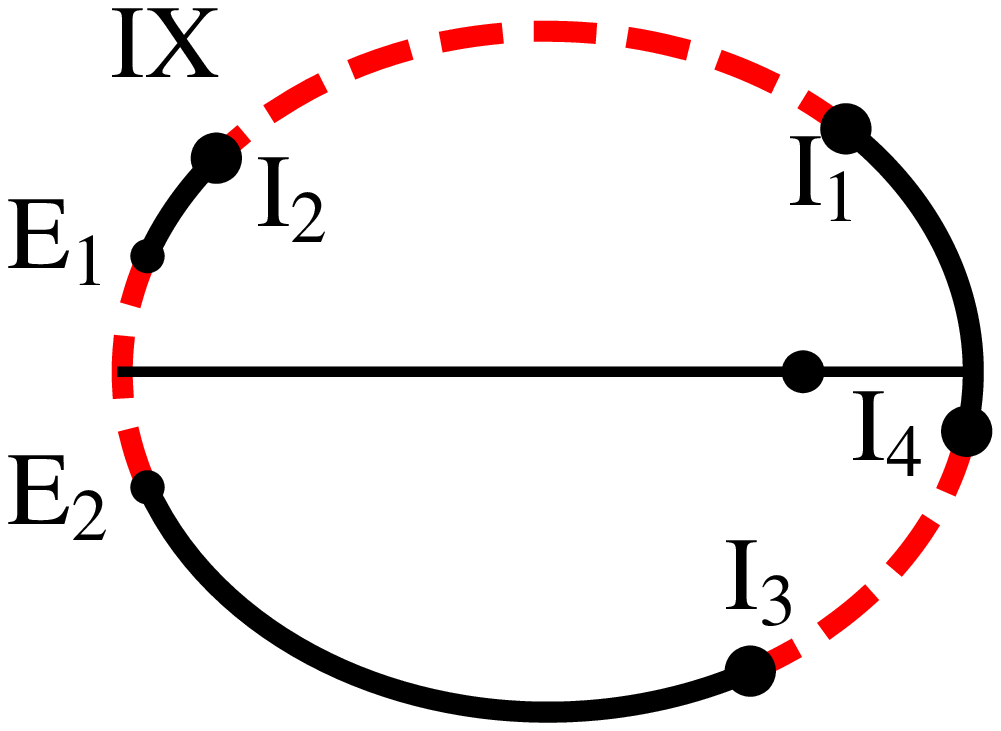}
  \end{center}
  \caption{Observable arc(s) of a TNO orbit that satisfy
  the distance and the inclination restrictions.
  Smaller points denoted with $E_1$ and $E_2$
  are intersection points of the orbit with the sphere $r=r_\text{max}$.
  Bigger points $I_1\dots I_4$ are intersection points of
  the orbit with the circles $b = \pm\epsilon$ on the sky.
  Solid and dashed arcs represent observable and non-observable
  parts of the orbit, respectively.
  }
  \label{fig:Ellipsen}
\end{figure}

Then we combine both observability constraints, from inclination
(Eq.~\ref{eq:phi}) and eccentricity (Eq.~\ref{eq:rmax}), into one, to find
the orbital arc (or arcs) that lie both in the observable latitudinal belt
and the observable sphere.
Typical geometries are sketched in Fig.~\ref{fig:Ellipsen}, assuming that
the pericenter is inside the observing latitudinal belt.
The intersection points of the orbit
with the visibility sphere $r = r_\mathrm{max}$ are denoted by $E_k$,
those with the visibility circle $|b| = \pm \epsilon$ by $I_k$
(indices $k$ increase with increasing true anomaly).
The point $E_1$ can lie before $I_1$,
between  $I_1$ and $I_2$,
or between $I_2$ and $I_3$,
giving three possibilities.
On the other hand, the point $E_2$ can reside
between  $I_2$ and $I_3$,
between $I_3$ and $I_4$,
or after $I_4$.
This yields $3\times 3 = 9$ possibilities in total, denoted by I...IX.
Remember that the position of the observable arcs in Fig.~\ref{fig:Ellipsen} 
corresponds to the case where the perihelion is inside the observable latitudinal belt.
If it is outside, the observable arcs change, giving rise to another set of nine 
possibilities.
Thus there are 18 possibilities in total.
Additionally, there are special cases.
One is $i < \epsilon$, where the entire orbit is inside the observable belt, so that 
the points $I_1\dots I_4$ do not exist.
Others are where the entire orbit is inside or outside the observable sphere,
so that the points $E_1$ and $E_2$ do not exist.

As an example, we take ellipse number III.
The object starts at the pericenter (where it is visible) and moves toward $I_1$.
Between $I_1$ and $I_2$, it is outside the observed latitudinal belt and
is invisible. Although it has a sufficiently
low ecliptic latitude up to $I_3$,
it is only detectable up to $E_1$,
because it gets too faint beyond that point.
Between $I_3$ and $I_4$ the object
is too far from from the ecliptic, and it stays outside the limiting sphere
until it reaches $E_2$. Starting from $E_2$, the
object is visible again.

Having found the observable arcs, we compute the fraction $f$ of the object's 
orbital  period it spends in these arcs for a given survey. If the survey had
a full ($360^\circ$) coverage of the ecliptic longitudes, that fraction
would directly give us the probability to detect the object. However, the
survey is confined to the longitude range with a certain width $\alpha$.
This width is often given in the papers, and where it is not, we
simply take $\alpha = \sqrt{\Omega} = 2\epsilon$,
where $\Omega$ is the observed sky area.
Thus the detection probability is $f$ multiplied by
$\alpha /360^\circ$.
This estimation assumes that all the observations are done within a period
of time that is much shorter than the orbital period of the TNO, so that its
proper motion can be neglected. This is the case for all the surveys we consider.
However, the same estimation assumes a uniform azimuthal distribution
of the TNOs. For plutinos, for instance, this is no longer valid, as they concentrate
preferably in two azimuthal zones ahead and behind Neptune's location.
Thus our algorithm may underestimate
the detection probability of resonant objects, at least in surveys targeted
at parts of the ecliptic where such objects are more numerous.

In this way, for each of the known TNOs,
we can calculate the detection probability in any survey.
We then calculate the probability $P_i$ that an object $i$
would be detected in any of the $N_\text{surveys}$ surveys:
\begin{equation}
	P_i = 1 - \prod_{k=1}^{N_\text{surveys}} \left(1-P_{ik}\right),
\label{eq:P}
\end{equation}
where $P_{ik}$ is the probability to detect an object $i$ in a survey $k$.
The advantage of equation (3) is that $1-P_{ik}$ gives the probability not to 
detect an object, ``shallower'' surveys make little contribution to the product and
thus to the total detection probability of very faint objects.
Therefore, it is deep surveys that dominate the result for faint objects.

Given the discovery probability $P_i$ of a given object,
we can augment the observed Kuiper belt to the ``true'' one
by counting that object $P_{i}^{-1}$ times.
In other words, we debias the observed Kuiper belt 
by setting the number of TNOs with the same orbital elements
as the known object to $P_{i}^{-1}$.

The number of surveys in Tab.~\ref{tab:observations} is $N_\text{surveys}=23$.
However, only nearly half of the 
$1260$ TNOs contained in the MPC database were found in 
these campaigns.
Another $\approx 600$ objects were discovered in other observations, some
serendipitously in surveys that did not aim to search for TNOs.
The circumstances of those observations have not been been published in all cases.
What is more, even for the campaigns listed in Tab.~\ref{tab:observations},
it is problematic to identify which particular set of $\approx 600$ 
objects out of $1260$ in total was found in those surveys.
Indeed, the papers that give a specific, identifiable list of newly discovered objects
(marked with an asterisk in  Tab.~\ref{tab:observations}) only cover $\approx 400$ TNOs.
We do not know under which
circumstances the remaining two-thirds of the TNOs were detected.
In other words, there is no guarantee that the parameters of those
unknown surveys ($m_{50}$, $\epsilon$ etc.) are similar to those listed in
Tab.~\ref{tab:observations}.
Furthermore, some of the surveys
in our list may not have reported their discoveries to the MPC.
As a result, it is difficult to judge how complete the MPC database is.
We can even suspect that there have been surveys not listed in
Tab.~\ref{tab:observations} that have
discovered TNOs not listed in the MPC.
Therefore, it does not appear possible to compile a complete version of 
Tab.~\ref{tab:observations} that would cover all known TNOs and
all discovery observations (together with their $\Omega$, $m_{50}$, and $\epsilon$).
Nor is it possible to get a complete list of all known TNOs together
with their orbital elements, along with information in which
particular survey each of the known TNOs was discovered.
To cope with these difficulties,
we make two assumptions.
First, we assume that the surveys listed in
Tab.~\ref{tab:observations}
are representative of all surveys that discovered TNOs.
Second, we assume that, conversely, the TNOs listed is the
MPC are respresentative of all the objects discovered in surveys
listed in Tab.~\ref{tab:observations}.
These two assumptions represent the main
shortcoming of our debiasing approach.

To check them at least partly and proceed with the debiasing,
we employed two different methods.
In the first method, we have randomly chosen $600$ TNOs out of the full list of known
objects and assumed that it is these objects that were discovered in the campaigns
listed in Tab.~\ref{tab:observations}.
We tried this several times for different
sets  of $600$ TNOs and found that the results (e.g., the elemental distributions 
and the total mass of the ``debiased EKB'') are  in close agreement.
In the second method, we have made an assumption that another set of $23$ 
similar surveys with similar detection success rate would have likely led to a discovery of 
all known TNOs. So we simply counted each survey twice and replaced Eq.~(\ref{eq:P}) by
\begin{equation}
	P_i = 1 - \prod_{k=1}^{N_\text{surveys}} \left(1-P_{ik}\right)^2.
\label{eq:better P}
\end{equation}
Again, the results turned out to be very close to those found with the first method.

\begin{figure}
\begin{center}
  \includegraphics[width=0.48\textwidth]{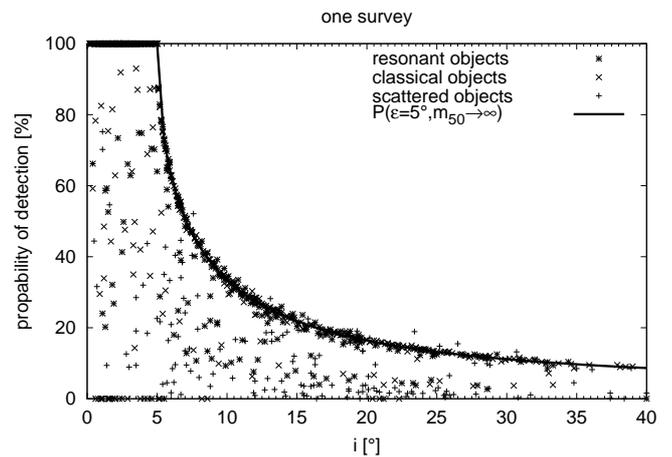}\\
\end{center}
\caption{
  The detection probability of all known TNOs (as function of their orbital
  inclinations) in a fiducial survey with a full coverage of a belt on the
  sky within $\epsilon = 5^\circ$ ecliptic latitude and the limiting magnitude
  of $m_{50} = 25\magn$. The curve is the formal detection probability of
  objects in circular orbits in the $m_{50}=\infty$ limit, but it
  approximates well the detection probability of many known TNOs in
  eccentric orbits in our fiducial survey. Objects which are below the curve are either those
  affected by the distance bias or have arguments of pericenters which are
  outside of our viewing field. Objects above the curve correspond to rare
  cases where the orbital eccentricity is high, aphelion lies in the
  observable belt, and the object is not too faint even near the aphelion.
  These are mostly scattered objects.
}
  \label{fig:Wahrscheinlichkeit}
\end{figure}

Figure~\ref{fig:Wahrscheinlichkeit} illustrates
the probabilities $P_{ik}$ to observe known TNOs
in a fiducial survey with $m_{50} = 25\magn$,
a latitudinal coverage of $\epsilon = 5^\circ$,
and a longitudinal coverage of $360^\circ$.
Let us start with an artificial case where all objects are in circular orbits.
If they were bright enough to be observed
(or equivalently, in the limiting case $m_{50} = \infty$),
they would all lie on the curve overplotted in Fig.~\ref{fig:Wahrscheinlichkeit}.
In particular, their detection probability would be 100\%
for $i < \epsilon$, and it would be $\epsilon/90^\circ = 5.6$\% for $i=90^\circ$.
If they are too faint for detection, their detection probability will be zero
regardless of the inclination.
The case of eccentric orbits is more complicated.
Then, the vast majority objects still are on the curve but, as seen in the figure,
there are many that lie below.
Either these are objects whose pericenter is outside of the latitudinal
belt $|b| < \epsilon$ or these are objects that cannot be observed over the entire 
orbits, even when they have sufficiently low ecliptic latitude,
because in some low-latitude parts of their orbits they are too faint to be visible.
In fact, a mixture of both cases is typical.
Finally, a few objects lie above the curve.
These are rare cases of objects in highly-eccentric orbits,
whose aphelia fall into the observable latitudinal belt,
and whose apocentric distances are not too large.
Such objects spend much of their orbital period near aphelia and are detectable there,
which raises their detection probability.

\begin{figure}
  \begin{center}
  \includegraphics[width=0.48\textwidth]{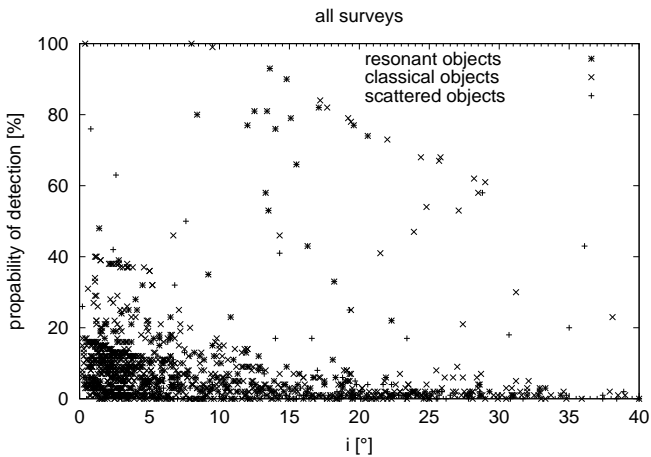}\\
  \end{center}
  \caption{
    The final detection probability of the known TNOs,
    calculated with Eq.~(\ref{eq:better P}).
    Included are all surveys from Table~\ref{tab:observations}.
  }
  \label{fig:Wahrscheinlichkeit2}
\end{figure}

Although the average detection probability in Fig.~\ref{fig:Wahrscheinlichkeit}
is quite high, this only holds for a complete coverage of the $|b| < \epsilon$ band 
on the sky.
In reality, only a limited range of the ecliptic longitude is covered.
The resulting detection probability $P_i$ of all known TNOs in all $23$ surveys,
calculated with Eq.~(\ref{eq:better P}) that takes into account actual
latitudinal coverage of the observational campaigns,
is plotted in Fig.~\ref{fig:Wahrscheinlichkeit2}.
Typical values are within $\sim 20$\% for near-ecliptic orbits and drop to a few percent for 
inclinations above $10^\circ$.

\subsection{Orbital element distributions of the Kuiper belt objects}

Having applied the debiasing procedure, we compared and analyzed 
the distributions of orbital elements of the known and the ``true''
EKB~--- separately for each class.

\begin{figure*}
  \begin{center}
  \includegraphics[width=0.98\textwidth]{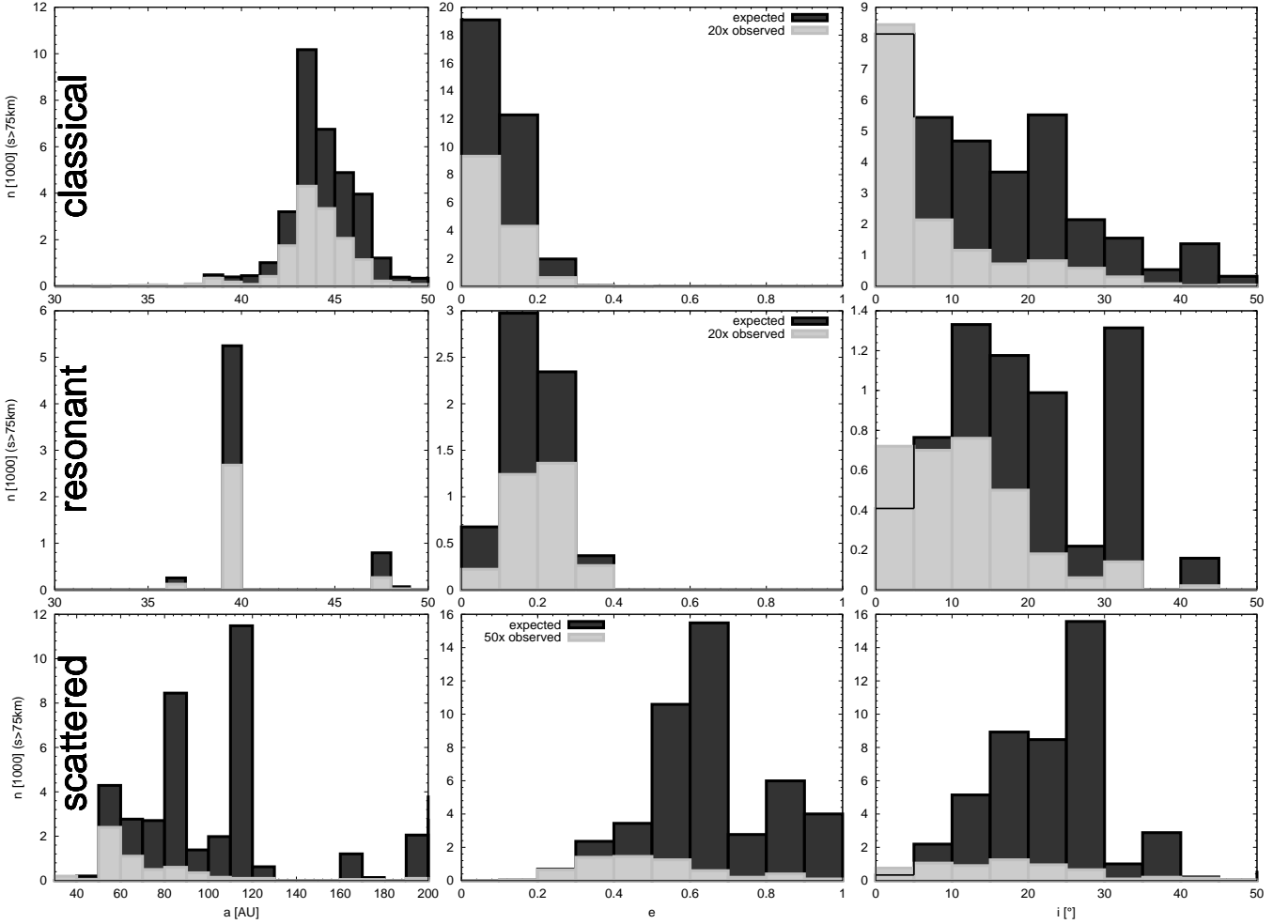}\\
  \end{center}
  \caption{
    Distribution of classical (top row), resonant (middle row),
    and scattered objects (bottom row), in terms of numbers of objects.
    Left column: semimajor axes, middle: eccentricities, right: inclinations.
    Black and grey bars in each panel represent the expected (unbiased) and observed
    populations, respectively.
    The numbers of the observed TNOs are magnified by 20 (classical and resonant objects)
    and 50 (scattered objects) for better visibility.
    Numbers are given in $1000$ for intervals with a width of $\Delta a = 1\AU$
    (classical and resonant), $\Delta a = 10\AU$ (scattered objects),
    $\Delta e = 0.1$ and $\Delta i = 5^\circ$ for all populations.
    }
  \label{fig:n_distr}
\end{figure*}

\begin{figure*}
  \begin{center}
  \includegraphics[width=0.98\textwidth]{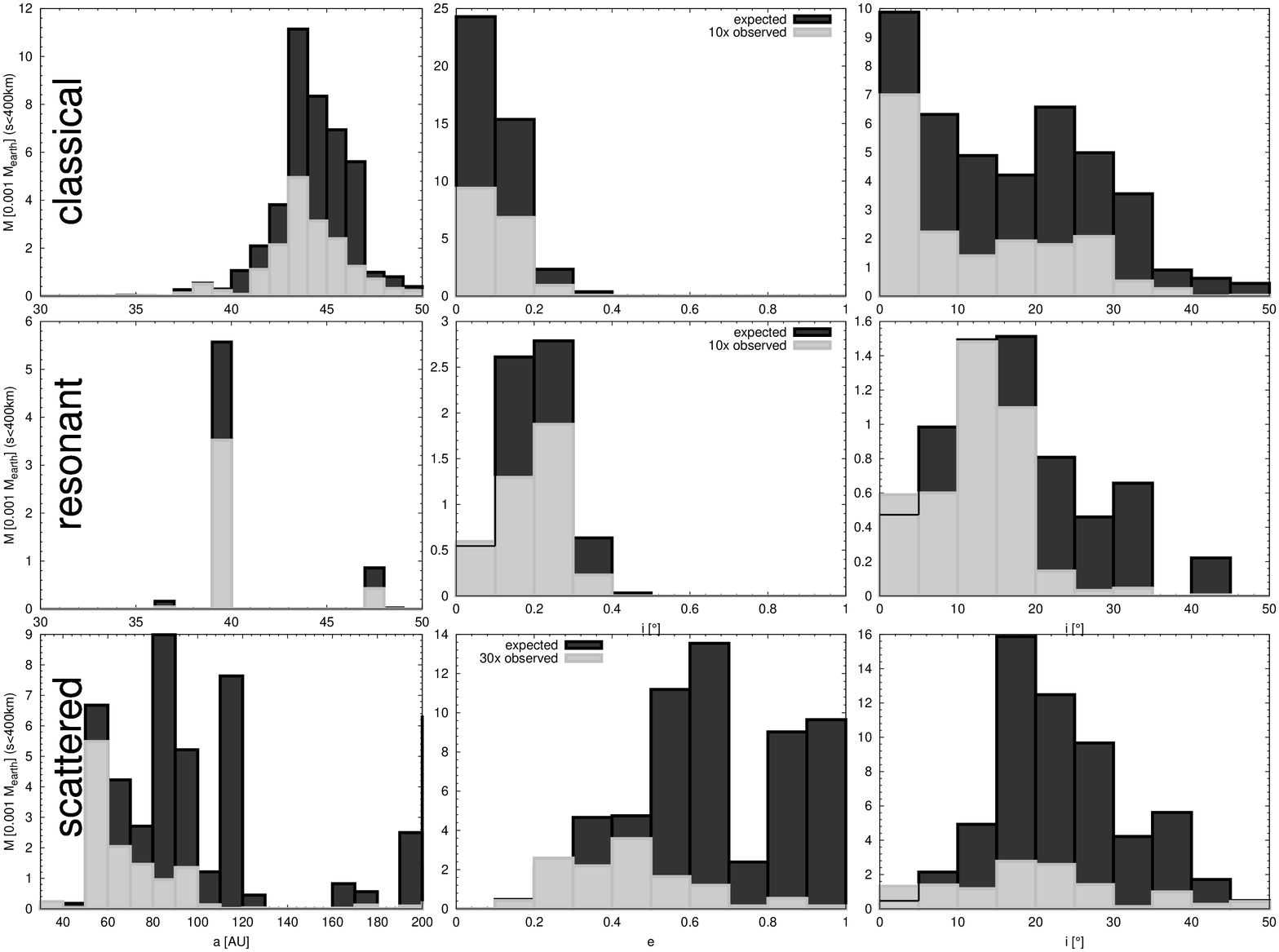}\\
  \end{center}
  \caption{
    Same as Fig.~\ref{fig:n_distr}, but in terms of mass contained in TNOs.
    }
  \label{fig:mass_distr}
\end{figure*}

Figures~\ref{fig:n_distr} and \ref{fig:mass_distr} show the distributions
in terms of numbers (for $s>75\km$) and
masses (for $s<400\km$) of objects per element's bin before and after debiasing.
The distribution in terms of numbers
depicted in Fig.~\ref{fig:n_distr} emphasizes smaller, more numerous, TNOs.
It is directly related to observational counts of TNOs and is also useful to alleviate 
comparison with similar work by the others.
In contrast, the distribution of TNO's mass in Fig.~\ref{fig:mass_distr} is dominated by
larger objects. It demonstrates more clearly where the wealth of the EKB {\em material}
is located, which aids placing the EKB in context of extrasolar debris disks.
Objects with $s<75\km$ were excluded from Fig.~\ref{fig:n_distr}, because
detections of the smallest objects are the least complete,
which would lead to a highly uncertain, distorted 
distribution.
And conversely, we excluded the biggest objects with $s>400\km$
from Fig~\ref{fig:mass_distr} to avoid large bin-to-bin variations stemming
from a few individual rogues. Failure to do this would lead, for example, to
a pronounced peak in the eccentricity distribution of resonant objects at $e =0.15 \dots 0.20$
produced by a single object, Pluto.

As seen in Figs.~\ref{fig:n_distr} and \ref{fig:mass_distr} for the classical Kuiper belt,
debiasing increases the total number and mass of objects, but the 
position of the maximum remains at $a\approx 44\AU$.
The same holds for the semimajor axis distribution of resonant objects, whose
peaks are preserved at known resonant locations.
In contrast, for the scattered objects, here are indications that a substantial unbiased
population  with larger semimajor axes of $80\AU \dots 120\AU$ might exist.
Some of them may be ``detached'' ($q > a_\mathrm{Neptune}$), while some others
may not (since the eccentricities of these TNOs are also large, see
middle panels in the bottom rows of Figs.~\ref{fig:n_distr} and \ref{fig:mass_distr}).
These conclusions should be taken with caution,
because the statistics of scattered objects is scarce and their debiasing factors are the largest.

The eccentricity distribution in Figs.~\ref{fig:n_distr} and \ref{fig:mass_distr}
shows moderate values $(e<0.2)$ for the
classical belt and reveals a broad maximum at $e \approx 0.1\dots 0.3$ for the resonant objects.
The maximum for the scattered objects appears to be located around $e\approx 0.5 \dots 0.6$.

As far as the inclination distribution (right panels in 
Figs.~\ref{fig:n_distr} and \ref{fig:mass_distr}) is concerned,
our analysis confirms the result by \citet{Brown-2001} who indentified two
distinct subpopulations in the classical Kuiper belt, a cold one with low
inclinations and a hot one with more inclined orbits.
The maxima of $0^\circ \dots 5^\circ$
and $20^\circ\dots 25^\circ$ that we found are consistent
with his results of ${2\fdg 6} ^{+0.2^\circ}_{-0.6^\circ}$ and
$17^\circ\pm 3^\circ$.

The inclination distribution of the resonant objects
reveals a broad maximum around $\approx 15^\circ$.
For comparison, \citet{Brown-2001} found a
maximum at ${10\fdg 2} ^{+2.5^\circ}_{-1.8^\circ}$.
A second maximum visible at $i=30^\circ \dots 35^\circ$
in the number distribution (Fig.~\ref{fig:n_distr}) is due to small objects
with a large debiasing factor, which are still
big enough not to fall under the $s<75\km$ criterion.
That is why in Fig.~\ref{fig:mass_distr} the same
peak is barely seen.

A clear difference between the number and mass distributions can be seen in the
bottom right panels of the two figures, too, which show the inclination distribution
of the scattered objects.
A large number of scattered TNOs can be found at $25^\circ \dots 30^\circ$ (Fig.~\ref{fig:n_distr}),
whereas their mass peaks at $15^\circ \dots 20^\circ$
(Fig.~\ref{fig:mass_distr}).
Interestingly, a recent paper by \cite{Gulbis-et-al-2010} yielded
${19\fdg 1} ^{+3.9^\circ}_{-3.6^\circ}$, which is
close to the maxima we find here.

\subsection{Albedos and sizes of the Kuiper belt objects}

\begin{figure}
  \begin{center}
  \includegraphics[width=0.48\textwidth]{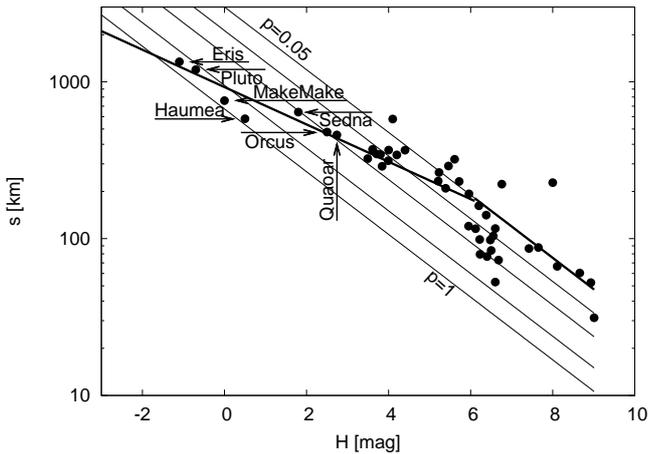}\\
  \end{center}
  \caption{
    Absolute magnitude -- radius relation for objects with known albedo.
    The biggest objects are labeled with their names.
    The thick solid line is a fit to this relation,
    Eqs.~(\ref{eq:s-H-fit1})--(\ref{eq:s-H-fit2}).
    Thin lines correspond
    to equal albedos of $p=0.05;\ 0.10;\ 0.20;\ 0.50;\ 1.00$.}
  \label{fig:albedo}
\end{figure}

To estimate the TNO sizes, we employed the V-band formula from
\citet{Kavelaars-et-al-2009}:
\begin{equation}
	H = m_\odot + 42.38 - 2.5\lg(4ps^2),
\end{equation}
where $m_\odot = -26.74\magn$ is the apparent V-magnitude of the Sun,
$p$ the albedo and $s$ the radius of an object in kilometers.
Solving for radius, we find
\begin{equation}
	s = 671.5\times \frac{10^{-0.2H}}{\sqrt{p}}\km.
\label{s_Hp}
\end{equation}
With this equation and albedo measurements from
\citet{Noll-et-al-2004,Stansberry-et-al-2008,Brucker-et-al-2009}, we calculated the radius of
objects with known albedo (Fig.~\ref{fig:albedo}).

Albedos inferred for a handful of big objects with $H \la 3$ turned out
to be high, which is indicative of a strongly reflecting surface material.
For instance, the surface of Haumea was found to be covered with $>92\%$
pure water ice \citep{Pinilla-Alonso-et-al-2009}.
\cite{Dumas-et-al-2007} reported for Eris $50\%$ methane ice on its surface
along with nitrogen and water ices, and ice tholin.
Smaller objects are coated with darker carbonaceous layers,
so their albedo is lower.
Note that objects between $6<H<7$ have a very strong
scatter, the reason for that being unknown. 
Albedos of the smallest TNOs with $7<H<9$ are typically close to $\approx 0.05$,
and there have been no measurements beyond $H=9$.
However, since the EKB is known to act as a reservoir of short-period comets,
we can use the measurements of cometary nuclei with sizes
of $\sim 1$--$10\km$ as a proxy for the reflectance properties of
the smallest TNOs. (The obvious caveat is that comets may have altered
their original surface properties as a result of their long residence
in the inner solar system.)
The typical albedo values of the nuclei
range from $0.02$ to $0.06$ \citep{Lamy-et-al-2004}.

On these grounds, to eliminate the dependence on albedo 
(which is not known for most of the Kuiper belt objects) from Eq.~(\ref{s_Hp}),
we have fitted the  sizes of the TNOs with known albedo
by an exponential function at $H < 6$
and assumed $p=0.05$ for all TNOs with $H \ge 6$.
This yielded a formula where $s$ is only a function of
$H$:
\begin{equation}
  s =  926\times 10^{-0.119H}\km \quad (H<6)
\label{eq:s-H-fit1}
\end{equation}
and
\begin{equation}
  s =  3000\times 10^{-0.2H}\km \quad (H\ge 6) .
\label{eq:s-H-fit2}
\end{equation}
The smallest object found so far is a scattered object with $H=15\magn$
which corresponds to a size of only $s = 3\km$. The smallest resonant object
has a radius of $s = 9.9\km$ $(H = 12.4\magn)$ and
the smallest classical one has $s = 12.5\km$.

\subsection{Mass of the Kuiper belt}

To translate the TNO sizes into mass requires assumptions for the bulk density.
We took the commonly used value of $\rho = 1 \gperccm$. 
This assumption is in accord with the values of $(0.6\dots 2.3)\gperccm$
found for a few individual TNOs \citep{Lacerda-2009}.
The resulting mass and number of objects in resonant, classical, and
scattered populations and in the entire Kuiper belt are listed in Tab.~\ref{tab:mass}.
The deduced ``true'' masses are several times higher
than in \citet{Fuentes-Holman-2008} who inferred $M_\text{CKB} = (0.008\pm 0.001)
M_\oplus$, $M_\text{SDO} = 0.010^{+0.021}_{-0.003} M_\oplus$, with a total
of $M_\text{tot} = 0.020^{+0.004}_{-0.003} M_\oplus$.
However, they considered the mass within $\pm 3^{\circ}$
around the ecliptic.
Since we investigated the full range of ecliptic latitudes, we deem the 
results consistent with each other.

\begin{table}
\centering
\begin{tabular}{|r|r|r|}
\hline
 					& before debiasing	& after debiasing\\
\hline\hline
$M_\text{CKB+RES}$ [$0.001 M_\oplus$]	& $4.4$			& $52$\\
$M_\text{SDO}$     [$0.001 M_\oplus$]	& $2.2$			& $65$\\
$M_\text{total}$   [$0.001 M_\oplus$]	& $6.6$			& $117$\\
\hline
$N_\text{CKB+RES}\ (s<75\km)$	& $1100$		&$39700$\\
$N_\text{SDO}\ (s<75\km)$	& $160$			&$45200$\\
$N_\text{total}\ (s<75\km)$	& $1260$		&$85000$\\
\hline
\end{tabular}
\caption{
  Masses and numbers of objects in the Kuiper belt.
  CKB is the classical Kuiper belt,
  RES are the resonant objects,
  and SDO denotes the scattered disk objects.
  }
\label{tab:mass}
\end{table}

One issue about the deduced mass of the entire EKB and its populations
is the influence of the uncertainties of the orbital elements inferred from
the observations. In many cases, the elements are known only roughly,
and some of them are not known at all. For $134$ out of $865$ known
classical objects, for instance, the observed arc was so short
that a circular orbit was assumed by the observers (these objects are
clearly visible as an $e=0$ stripe in the upper panel of Fig.~\ref{fig:a-e-i-dist}).
How could a change in the orbital elements of an object affect the debiasing procedure
and the final estimates of the parameters of the ``true'' EKB?
Obviously, if a true value of {\em one} of the three elements of a TNO
($a$, $i$, or the absolute magnitude $H$) is larger than the one given in the 
database, the detection
probability will be overestimated and the estimated number of similar objects
in the ``true'' EKB underestimated. The eccentricity plays a special role in this case.
Increasing it would not automatically lead to a lower detection probability,
the pericenter distance decreases while the apocenter increases, so that the total
detection probability depends also on $a$. However, a combined variation of two
or more elements may alter the results in either direction.
As an example, let us consider a scattered object with $a = 1057\AU$ and $e=0.977$,
which has a pericenter distance of $q = 24.3\AU$.
Decreasing, for instance, both $a$ and $e$ by $5\%$ would lead to a 
pericenter at $q = 72\AU$, which would result in a significantly lower detection 
probability and therefore in a higher contribution of that object to the 
estimated total mass. In contrast, we may consider a classical object with $a=40\AU$ and $e=0.2$,
which cannot be observed near the apocenter. Again, decreasing both values by $5\%$
would now reduce the aphelion distance to detectable values, so that the
detection probability would increase.

From published observational results, we assume $5$--$10\%$ as a
typical error for the orbital elements.
To quantify possible effects, we used the following Monte-Carlo procedure.
We assumed that the orbital elements and the absolute magnitude $\{a,e,i,H\}$ 
are known with a certain relative accuracy $\sigma$ (for simplicity, the same for all 
four elements). Then, we randomly generated 
$\{a,e,i,H\}$ -sets for each of the known TNOs assuming that each element
of each object is uniformly distributed between $x - \sigma$ and $x+ \sigma$,
where $x$ is the cataloged value. For this hypothetical EKB, the debiasing procedure
was applied and the expected masses and number of objects in the ``true'' Kuiper belt
were evaluated. This procedure was repeated $10,000$ times (for $10,000$ realizations
of the observed Kuiper belt, that is to say).
From these calculations we excluded the scattered objects,
because varying their orbital elements would lead to
extremely low detection probabilities.
The results for several $\sigma$ values between $5\%$ and $15\%$ are
listed in Table~\ref{tab:errors}.
It is seen that the effect of the uncertainties of the orbital elements on
the global parameters of the ``true'' EKB is moderate except for the SDOs.
Since the SDOs have ``extreme'' orbital elements, compared to the classical belt,
an uncertainty of, e.g., $25\%$ would alter the total mass by a factor of $2$.

Interestingly, the net effect of the increasing $\sigma$ is that the mean TNO detection
probabilities slightly decrease, which leads to somewhat higher
estimates for the mass of the EKB populations and the whole Kuiper belt.

\begin{table}
 \centering
\begin{tabular}{|c|c|c|c|}
\hline
$\sigma [\%]$ & $M_\text{CKB}$ & $M_\text{RES}$ & $M_\text{CKB+RES}$ \\
\hline\hline
$5$  & $46.1 \pm 1.6$ & $8.7 \pm 0.6$ & $54.8 \pm 1.7$   \\
$10$ & $48.3 \pm 2.2$ & $9.1 \pm 0.7$ & $57.3 \pm 2.3$   \\
$15$ & $53.3 \pm 3.5$ & $9.7 \pm 0.8$ & $63.0 \pm 3.6$  \\

\hline
\end{tabular}
\caption{
  Masses of objects in the Kuiper belt, as a function of
  the assumed relative accuracy $\sigma$, with which orbital elements
  of TNOs were deduced from observations.
  Abbreviations and units are as in Tab.~\ref{tab:mass}.
}
\label{tab:errors}
\end{table}

\subsection{Size distribution of the Kuiper belt objects}

We now come to the size distribution of KBOs.
The exponents $q$ of the differential size distribution
$N(s)\total s \propto  s^{-q}\total s$ after debiasing were 
derived 
with the size-magnitude relation (\ref{eq:s-H-fit1})--(\ref{eq:s-H-fit2}).
In doing so, we have chosen the size range $50\km < s < 170\km$ $(8.9>H>6)$,
and we determined the size distribution index separately for different
populations of TNOs and their combinations.
For the CKB, the result is $q = 4.3\pm 0.2$.
The resonant objects reveal a steeper slope of $5.1\pm 0.1$,
with plutinos (in 3:2 resonance with Neptune) having $5.3\pm 0.1$ 
and twotinos (2:1 resonance) having $4.0\pm 0.1$.
This results in $4.4\pm 0.2$ for classical and all resonant objects together.
In contrast, the scattered objects have $2.8\pm 0.1$.
Altogether, we find $3.6\pm 0.1$ for the entire EKB (classical, resonant, and scattered TNOs).

Our results are largely consistent with previous determinations
(Table~\ref{tab:sizeindex}). For the CKB, for instance,
the range between $3.6\pm 0.1$ \citep{Chiang-Brown-1999} 
and $4.8^{+0.5}_{-0.6}$ \citep{Gladman-et-al-1998} was reported.
In this comparison, one has to take into account that different authors
dealt with somewhat different size intervals.
\cite{Chiang-Brown-1999} considered objects between $(50\dots 500)\km$,
\cite{Gladman-et-al-2001} and \cite{Trujillo-et-al-2001a} between $(50\dots 1000)\km$,
and \cite{Donnison-2006} between $(120\dots 540)\km$ $(7>H>2)$.
For the SDOs, our results are also consistent within the error bars 
with \cite{Donnison-2006}. However, for the resonant objects our result departs from his
appreciably.

\begin{table}[ht!]
\centering
\begin{tabular}{|c|c|c|c|}
\hline
CKB &	RES	& SDO & reference\\
\hline\hline
$4.3\pm 0.2$		&  $5.1\pm 0.1$	& $2.8\pm 0.1$	& this paper\\
$4.8^{+0.5}_{-0.6}$	&		&		& \cite{Gladman-et-al-1998}\\
$4.0\pm 0.5$		&		&		& \cite{Jewitt-et-al-1998}\\
$3.7\pm 0.2$		&		&		& \cite{Luu-Jewitt-1998}\\
$3.6\pm 0.1$		&		&		& \cite{Chiang-Brown-1999}\\
$4.4\pm 0.3$		&		&		& \cite{Gladman-et-al-2001}\\
$4.0^{+0.6}_{-0.5}$     &		&		& \cite{Trujillo-et-al-2001a}\\
$4.05 \pm 0.2$		&		&		& \cite{Bernstein-et-al-2004}\\
$3.97\pm 0.15$		&$3.30\pm 0.37$ &$3.02\pm 0.32$	& \cite{Donnison-2006}\\
\hline
\end{tabular}
\caption{Size distribution index of the Kuiper belt populations.}
\label{tab:sizeindex}
\end{table}

\begin{figure}
  \begin{center}
  \includegraphics[width=0.48\textwidth]{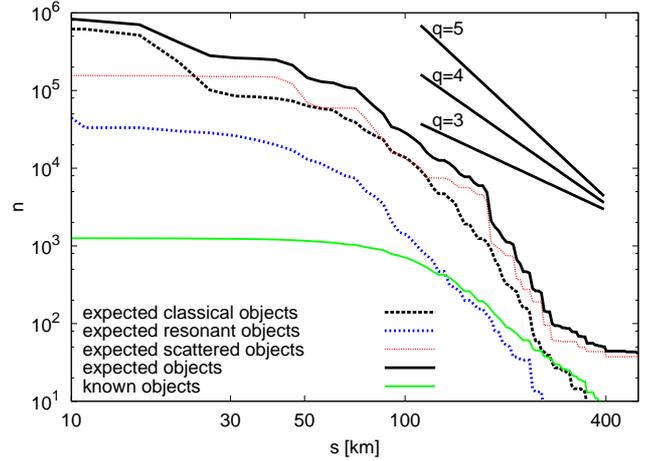}\\
  \end{center}
  \caption{
    Cumulative numbers of the expected and known Kuiper belt objects with a $\Delta s = 5\km$ resolution.
    For an easier orientation, straight lines show the slopes that would
    correspond to the differential size distribution indexes of $q=3$, $4$, and $5$.
  }
  \label{fig:groesse}
\end{figure}

Figure \ref{fig:groesse} shows cumulative numbers of the expected Kuiper belt objects
larger than a given size.
In agreement with \citet{Donnison-2006}, the profile flattens for objects
$s\lesssim 60\km$ ($H<7$).
The break in the size distribution at radii of several tens of kilometers
reported by some authors, e.g., at $s\approx 30\km$ by 
\citet{Bernstein-et-al-2004} and \citet{Fraser-2009}
can neither be clearly identified nor ruled out with our
debiasing algorithm.

%------------------------------------------------------------------------------------------------------------

\section{Dust in the Kuiper belt}

\subsection{Simulations}

We now move on from the observable ``macroscopic'' objects in the EKB to
the expected debris dust in the transneptunian region.
We employ the technique to follow the size and radial distribution of solids 
(from planetesimals down to dust)
in rotationally-symmetric debris disks, developed in previous
papers \citep{Krivov-et-al-2000,Krivov-et-al-2005,Krivov-et-al-2006,%
Krivov-et-al-2008,Loehne-et-al-2008,Mueller-et-al-2009}.
Our numerical code,  {\it ACE} ({\it Analysis of Collisional Evolution}),   
solves the Boltzmann-Smoluchowski kinetic equation over a grid of masses $m$,
periastron distances $q$, and orbital eccentricities $e$ of solids.
It includes the effects of stellar gravity, direct radiation pressure,
as well as disruptive and erosive collisions.
Gravitational effects of planets in the system are not simulated with {\it ACE}
directly. Since we do expect Neptune to affect the dust disk in the EKB region
in several ways, we will discuss this later in Sect.~5.
The code outputs, among other quantities, the size and radial
distribution of disk solids over a broad size range from sub-micrometers
to hundreds of kilometers at different time steps, and the code is fast
enough to evolve the distribution over gigayears.

\begin{figure}
  \begin{center}
  \includegraphics[width=0.48\textwidth]{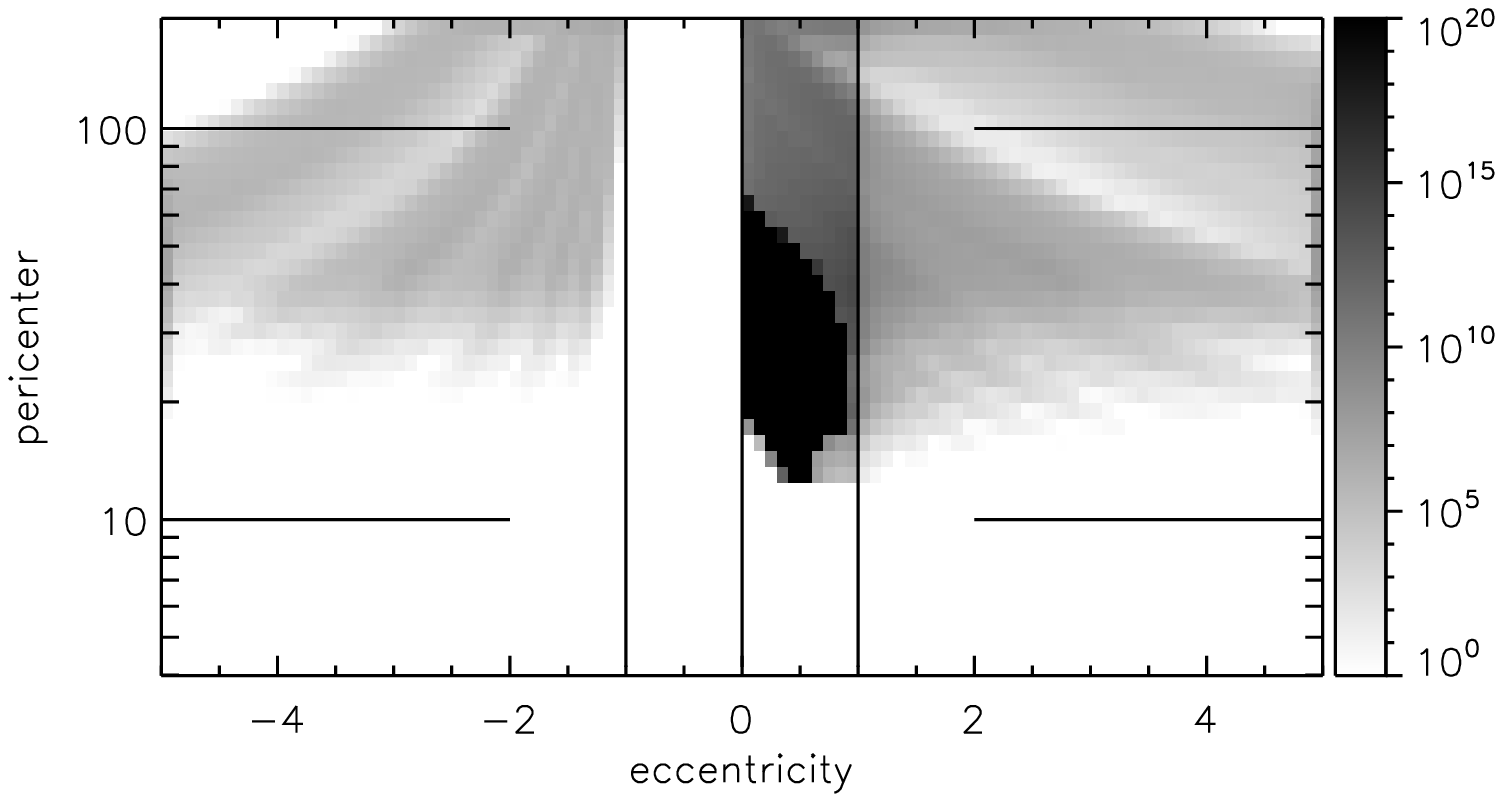}
  \includegraphics[width=0.48\textwidth]{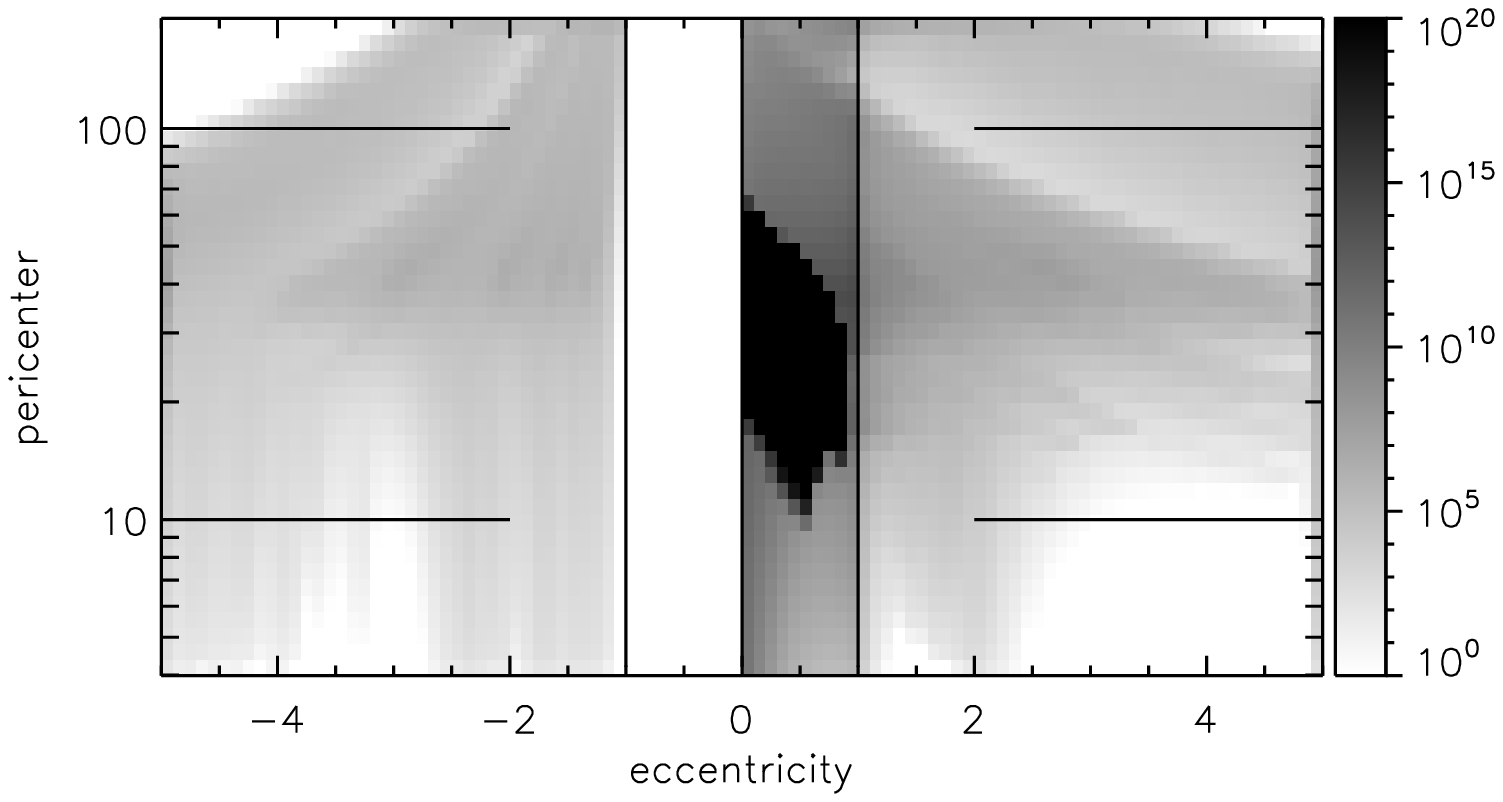}\\
  \end{center}
  \caption{
    Phase-space distribution (in $e,q$-plane) of dust maintained by 
    the expected EKB.
    Top: without P-R effect,
    bottom: with P-R effect.
    Pericenter distances are in AU. 
    Eccentricities $e<-1$ correspond to anomalous hyperbolas open outward from
    the Sun.
    Grayscale gives the total cross section of dust, integrated over
    all sizes (in arbitrary units).
    }
  \label{fig:phase-space}
\end{figure}

As explained in Sect.~1,
an effect of particular importance in transport-dominated disks is the
Poynting-Robertson drag.
The latter is now implemented in {\it ACE} through an appropriate diffusion
term in the space of orbital elements, coming from the classical
orbit-averaged equations for $\dot{q}$ and $\dot{e}$ \citep[e.g.][]{Burns-et-al-1979}.
To clearly see the role played by the P-R drag,
we ran {\it ACE} twice, with and without P-R.
Figure \ref{fig:phase-space} illustrates the distribution of the perihelion distance $q$
versus orbital eccentricity $e$ of dust grains.
This distribution was calculated both without (top panel) and with P-R effect (bottom panel).
The location of the main belt is recognizable as a dark grey region
in each of the plots.
Clearly visible is the dual role played by the P-R drag.
On the one hand, it lowers the pericentic distances of dust particles,
filling the lower parts of the bottom panel. On the other hand, it circularizes
the orbits of dust particles. This can be seen, for example, as a slight concentration
of particles, whose  pericenters are located inside the main belt, towards the $e=0$ line.

To model the collisional evolution of the EKB, as well as the distribution and
thermal emission of the EKB dust, we have to assume certain optical and mechanical
properties of dust. This, in turn, necessitates assumptions about its
chemical composition. The surface composition of a number of bright EKB objects
has been measured \citep[see, e.g.,][for a recent review]{Barucci-et-al-2008};
see also discussion in Sect.~2.5.
These objects turned out to have surfaces with very different 
colors and spectral reflectances. Some objects show no diagnostic
spectral bands, while others have spectra showing signatures of various
ices (such as water, methane, methanol, and nitrogen). The diversity in
the spectra suggests that these objects represent a substantial range of
original bulk compositions, including ices, silicates, and organic
solids. A single standard composition that could be adopted to represent
``typical'' EKB dust grains is therefore difficult to find.
For the sake of simplicity, for the collisional simulation with {\it ACE}
we choose an ideal material with $\rho = 1\gperccm$ and geometric
optics, leading to the radiation pressure efficiency $Q_{pr}=1$.
Next, an important property that we need for the collisional simulations 
is the critical specific energy $Q_D^*$, which is the ratio
of impact energy and mass of the target. It is calculated by the sum of two power laws
\citep[][and references therein]{Krivov-et-al-2005,Loehne-et-al-2008}
\begin{equation}
	Q_D^* = A_s \left(\frac{s}{1\m}\right)^{3b_s} + A_g \left(\frac{s}{1\km}\right)^{3b_g} ,
\end{equation}
where the first and the second term represent the strength and the gravity regime,
respectively.
We took values thought to be typical of low-temperature ice:
$A_s = 10^6 \ergperg$,
$A_g = 2\times 10^6 \ergperg$,
$3b_s = -0.37$ and $3b_g = 1.38$.

In the collisional simulation with {\it ACE}, we refrain from testing various
possible material compositions of dust, for the following reasons.
First, each {\it ACE} run requires up
to a few weeks of computing time in parallel mode on 8-16 kernels.
Second, we would need to consistently modify both the optical constants
of the assumed material (that determine the strength of the radiation pressure
force through $Q_{pr}$) {\em and} its mechanical properties that control
the collisional cascade. The latter would be problematic because of the lack of
experimental data, for instance on $Q_D^*$, for non-icy materials.
Nevertheless, in Sect. 4, we will test the influence of various materials
on the resulting termal emission of the EKB dust to get a rough idea of
the material dependence of the simulation results.

In the {\it ACE} simulations, we used the following
size--pericentric distance--eccentricity mesh.
The minimum grain radius was set to $0.1\mum$ 
and the variable mass ratio in the adjacent bins between $4$ (for largest TNOs)
and $2.1$ (for dust sizes).
The pericenter distance grid covered 41 logarithmically-spaced values
from 4~AU to 200~AU.
The eccentricity grid contained 50 linearly-spaced values between
$-5.0$ and $5.0$ (eccentricities are negative in the case of smallest
grains with $\beta>1$, whose orbits are anomalous hyperbolas, open
outward from the star).
The distance grid used by {\it ACE} to output distance-dependent
quantities such as the size distribution was $100$ values
between 4~AU and 400~AU.

In many previous studies, the initial radial and size distributions of dust parent 
bodies~---  planetesimals~--- were taken in the form of power laws, 
with normalization factors and indices being parameters of the simulations.
In this paper, we use a different approach. To take advantage of our knowledge
of the (largest) parent bodies, TNOs, we directly filled the $(m,q,e)$-bins 
at the beginning of each simulation with the objects of the ``true'' Kuiper belt.
For comparison, we also made a run, where we populated the bins with
known TNOs only (without debiasing).

As already described, our ``true'' distribution contains only big objects
with radii greater than $\sim 5\km$, whereas in reality small objects at all 
sizes down to dust must be present, too.
If we started a simulation without smaller objects, the collisional cascade would 
take many gigayears to produce a noteworthy amount of dust and to reach collisional
equilibrium.
Accordingly, we have extrapolated the contents of the filled bins towards smaller
sizes with a slope of
$q=3.03$ for objects between $100\m<s<75\km$ (in the gravity regime)
and $q=3.66$ for objects smaller than $100\m$  (in the strength regime),
following \citet{Obrien-Greenberg-2003}. 
Note that the adopted slope in the gravity regime is roughly consistent
with Fig. \ref{fig:groesse}.
The break at radii of several tens of kilometers
reported by \citet{Bernstein-et-al-2004} and \citet{Fraser-2009}
was not included.
In the course of the collisional evolution,
this artificial distribution corrects itself until it
comes to a collisional quasi-steady state.
The latter is assumed to have been reached, when the size distribution
no longer changes its shape and
just gradually moves down as a whole as a result of collisional depletion
of parent bodies \citep{Loehne-et-al-2008}.
We find that after $\la 100\Myr$ 
a collisional quasi-steady state sets in for all solids in the strength regime 
(i.e., smaller than $\sim 100\m$).
This particularly means that the system ``does not remember'' anymore
the assumed initial distribution in this size range.

We stress that the extrapolation of the observable EKB toward smaller sizes
described above should not be misinterpreted as an attempt to describe the primordial 
size distribution of solids in the early EKB. The latter is set by the mechanism
of the initial planetesimal accretion, which is as yet unknown. In the
standard scenarios of ``slow'' accretion \citep[e.g.][]{Kenyon-Bromley-2008}
a broad size distribution is expected. In contrast, ``rapid'' scenarios, such as
the ``primary accretion'' mechanism proposed by \citet{Cuzzi-et-al-2007}
or ``graviturbulent'' formation triggered by transient zones of high pressure 
\citep{Johansen-et-al-2006} or by streaming instabilities
\citep{Johansen-et-al-2007}  all imply that most of the mass of just-formed 
planetesimals was contained in $s \sim 100\km$ bodies.
Whatever mechanism was at work, and whatever size distribution the 
EKB in the early solar system had, in this study we are only interested in the
present-day EKB. Thus the purpose of the extrapolation described above is
merely to choose the initial size distribution across the sizes 
that would be as close to collisional equilibrium with the present-day EKB
as possible.

\subsection{Size distribution of dust}

\begin{figure}
  \begin{center}
  \includegraphics[width=0.48\textwidth]{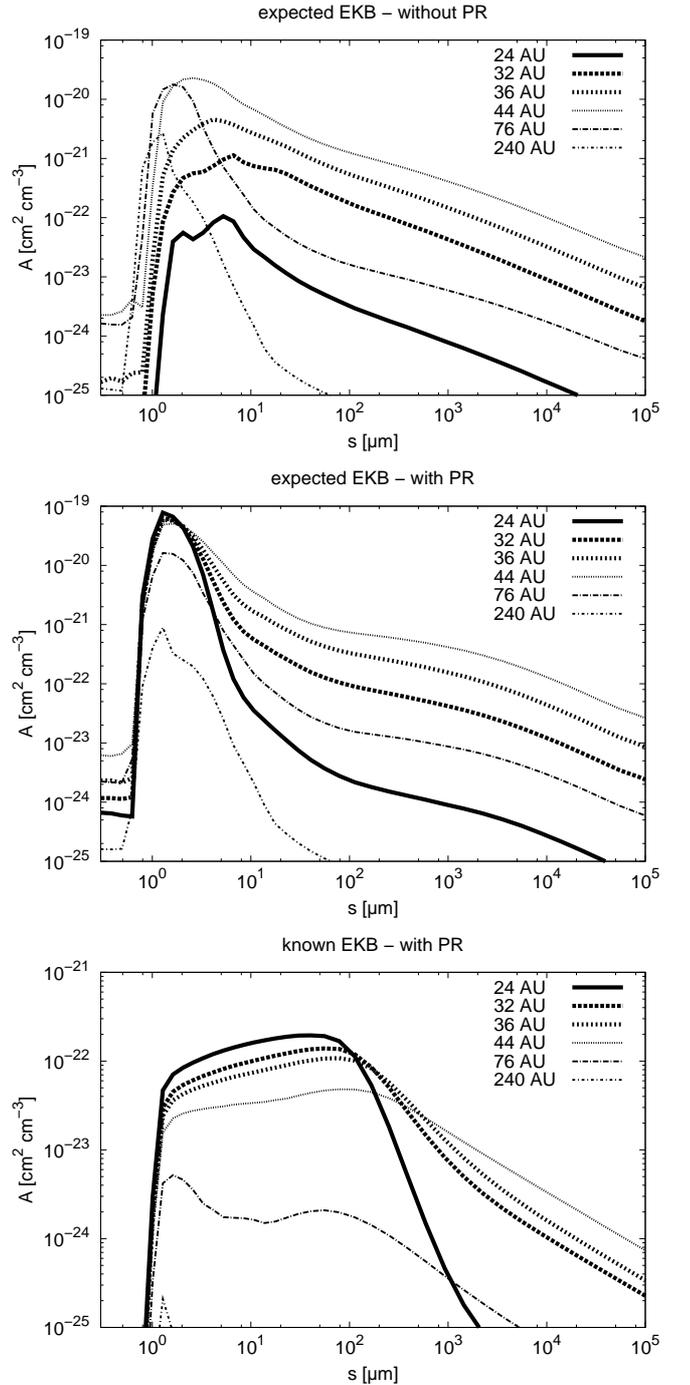}
  \end{center}
  \caption{Size distribution of the Kuiper belt dust at different distances.
  The vertical axis gives the cross section density per size decade.
  Top: the debiased EKB, without P-R effect;
  middle: the debiased EKB, with P-R effect;
  bottom: known EKB objects only, with P-R effect.
}
  \label{fig:size}
\end{figure}

Figure~\ref{fig:size} depicts the simulated size distribution of the EKB dust.
We present three cases:
for the debiased EKB without (top panel) and with P-R included 
(middle), as well as for the known EKB objects with P-R effect, for comparison 
(bottom).
To explain the gross features of the size distributions shown in 
Fig.~\ref{fig:size},
we introduce the ratio of radiation pressure to gravity,
usually denoted as $\beta$ \citep{Burns-et-al-1979}.
If a small dust grain is released after a collision from a nearly circular orbit,
its eccentricity is $e \approx \beta (1-\beta)^{-1}\propto s^{-1}$.
This implies higher eccentricities for smaller grain sizes.
The orbits of sufficiently small grains with $\beta$ exceeding $\sim 0.5$
are unbound. Accordingly, the grain radius that corresponds
to $\beta = 0.5$ is commonly referred to as blowout limit.
The blowout size for the assumed ideal material in the solar system is
$s_\mathrm{blow} \approx 1.2\mum$.
Typically, the amount of blowout grains instantaneously present in the
steady-state system is much less than the amount of slightly larger
grains in loosely bound orbits around the star.
This is because the dust production of the grains of adjacent sizes is
comparable, but the lifetime of bound grains (due to collisions) is much
longer than  the lifetime of blowout grains (disk-crossing timescale). 
This explains a drop in the size distribution around the blowout size
which is seen in all three panels of Fig.~\ref{fig:size}.

Another generic feature of the size distribution is that it becomes narrower
at larger distances from the Sun. Were the parent bodies all confined
to a narrow radial belt, the distribution far outside would appear as a 
narrow peak composed only of small, high-$\beta$,
barely bound grains sent by radiation pressure into eccentric orbits
with large apocentric distances.
However, in the case of the EKB this
effect is somewhat washed out, since the radial distribution of parent bodies
themselves (mostly, of scattered objects) is extended radially, as
discussed in Sect. 3.3. As a result, the size distribution even at relatively
large distances  (e.g. $76\AU$) is a superposition of such a narrow distribution
and a background broad distribution of particles produced at those distances directly.
Only at largest distances, at which hardly any parent bodies are present
(see $240\AU$ curve), the size distribution transforms to a predicted narrow
peak adjacent to the blowout size.

A direct comparison of our two simulations for the debiased EKB, without and
with P-R effect, reveals some differences.
One obvious~--- and expected~---  difference is the one between the $24\AU$ curves.
At this distance (and all the others inside the main belt) parent bodies
are nearly absent; there are only some scattered TNOs,
see Fig.~\ref{fig:a-e-i-dist}.
Accordingly, without P-R nearly no dust is present there. However, a substantial
amount of small particles is present there in the P-R case, because these are transported
there by the P-R drag.

Outside $\sim 30\AU$, the size distributions without and with P-R
show more similarities than dissimilarities.
In particular, the maximum of the cross section
at $s=2\mum$ is nearly the same.
At sizes $s \ga 1\mm$, the curves roughly follow a classical Dohnanyi's law
(cross section per size decade $\propto s^{-0.5}$).
The main difference is a dip of the size distribution in the region of the classical
EKB that occurs at sizes of $s=100\mum$ in the P-R case, which is easy to explain.
The $100\mum$ grains in the classical EKB region stay in nearly-circular orbits,
because their $\beta$ ratio is small and radiation pressure-induced eccentricities are low.
These grains are mainly destroyed in collisions with most abundant smaller
grains, several $\mum$ in size.
In the non-P-R case, the latter grains have their pericenters
within the classical belt. Thus the collisions are ``grazing'', the collisional
velocities relatively low, and the collisional desctruction of $100\mum$ grains
relatively inefficient.
When the P-R effect is switched on, this changes. The P-R transport lowers the
pericenters of smaller projectiles, and in the classical belt, they collide with $100\mum$
grains at higher speeds, which enhances their destruction and produces the dip.
Note that this effect is absent farther out from the Sun. At a $76\AU$ distance,
for example, the collisions between smaller grains and $100\mum$ particles
always occur far from the pericenter. Thus the P-R effect has little influence
on the collisional velocities, and the size distribution
in the non-P-R and P-R cases is similar.

\begin{figure*}
  \begin{center}
  \includegraphics[width=0.99\textwidth]{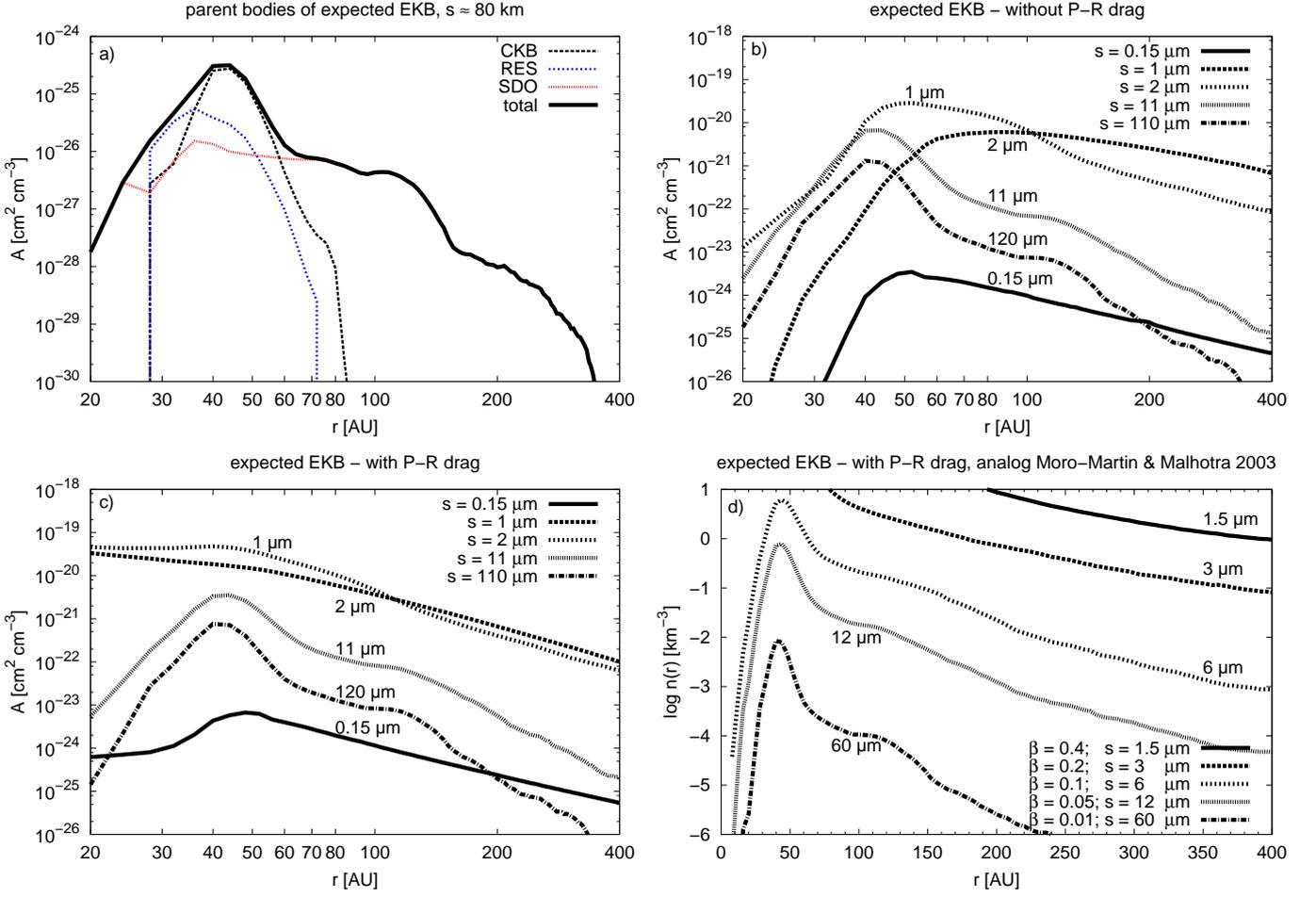}\\
  \end{center}
  \caption{
     Radial distribution of parent bodies ((a), from $64\km$ to $100\km$ in radius)
     and different-sized dust grains
     without (b) and with P-R drag (c).
     The vertical axis is the cross section density.
     The d panel is the same as c, but replotted in the same 
     units as Fig.~1a in \citet{MoroMartin-Malhotra-2003}.
   }
  \label{fig:size_different_distance}
\end{figure*}

However, the dust distributions computed with P-R effect,
but for expected (i.e., debiased) EKB and known (not debiased) EKB, which
are shown in middle and bottom panels of Fig.~\ref{fig:size}, exhibit a striking difference.
The P-R effect has only a moderate influence on dust produced by the
expected (debiased) EKB, but a strong one on dust generated solely by known TNOs.
This needs to be explained.
The debiasing procedure makes the EKB more densely populated, and the resulting
increase in the dustiness shortens collision timescales to make them
comparable with the P-R transport timescales.
The resulting optical depth of the dust disk is such that it lies
roughly between the collision-dominated and transport-dominated regimes.
Without debiasing the parent body population, the dustiness of the disk
is by two orders of magnitude lower, and so is the optical depth of the dust disk.
At that optical depth level, the EKB dust disk would be transport-dominated
below $s \la 100\mum$ (but still collision-dominated at larger sizes).
This is illustrated by the lowest panel in Fig.~\ref{fig:size} that presents
the size distribution of dust that would be produced by known TNOs.
It is seen that the size distribution in such a transport-dominated disk
differs from that in a collision-dominated one qualitatively.
From $s \sim 100\mum$ down to blowout limit, the size distribution flattens and turns 
over. This is because the smaller the grains, the faster their inward P-R drift.
This transport removes small grains from the collisionally active region
and thus they are present in smaller amounts.
As a result, the maximum of the cross section density shifts towards 
$s \sim 100\mum$ particles.

\subsection{Radial distribution of dust}

Figure~\ref{fig:size_different_distance} presents the radial distribution
of dust parent bodies and their dust, the latter simulated without and with
 P-R transport.

We start with a radial distribution of parent bodies, TNOs themselves,
shown in Fig.~\ref{fig:size_different_distance}a.
In contrast to Fig.~\ref{fig:mass_distr}, we plot here the total cross section
of the TNOs instead of the mass they carry, because it is the cross section
that characterizes the efficiency of TNOs as dust producers.
Besides, we use the distance from the Sun instead of the semimajor axis as an argument.
Specifically, we plot the cross section in the $80\km$-sized TNOs, but the
radial profile for larger objects look similar.
As expected, the distribution peaks in the region of the main belt ($40$--$50\AU$),
where about 90\% of the cross section comes from the classical EKB
objects.
Outside $\sim 60\AU$, the cross section is solely due to scattered objects.
The distribution of the latter is quite extended radially, it is nearly flat
over a wide distance range from $\approx 35\AU$ to more than $100\AU$.

We now move to a discussion of the radial distribution of dust.
As noted above, smaller grains with higher $\beta$ ratios acquire
higher orbital eccentricities.
As the eccentricities of particles slightly above the blowout limit are the highest,
their radial distribution is the broadest, whereas larger particles
stay more confined to their birth regions.
In Fig.~\ref{fig:size_different_distance}b
this effect can be seen from how the curves gradually change from the largest ($s=110\mum$)
to the smallest bound grains ($2\mum$).
The former essentially follow the distribution of the parent bodies,
while the latter exhibit a more extended, flatter radial profile.
Finally, blowout grains (e.g., those with $s = 0.15\mum$ have an $\propto r^{-2}$
distribution, as expected for a set of hyperbolic orbits streaming outward from their 
birth locations.

Including the P-R drag (Fig.~\ref{fig:size_different_distance}c) does little with largest grains, 
but modifies the profile of smaller ones ($\sim 1\mum$) substantially.
The P-R transport inward steepens their profile.
Besides, small particles are now present in high amounts at smaller 
distances, even where no parent bodies are present, in contrast to the case
without P-R.

Figure~\ref{fig:size_different_distance}d is the same as
Fig.~\ref{fig:size_different_distance}c,
but replotted in the same units as Fig.~1a in \citet[][]{MoroMartin-Malhotra-2003}.
Instead of treating the actual distribution of TNOs and their collisional dust 
production, they assumed a simplified narrow birth ring between $35\AU < a < 50\AU$ and 
a constant dust production rate. Nonetheless, the general behavior remains 
the same.
At $r\approx 50\AU$ the number density begins to decrease rapidly
but we still have some parent bodies outside $50\AU$
which produce dust, so that in our case the number density decreases more gently.

\subsection{Coupled size-radial distribution}

\begin{figure}
  \begin{center}
  \includegraphics[width=0.48\textwidth]{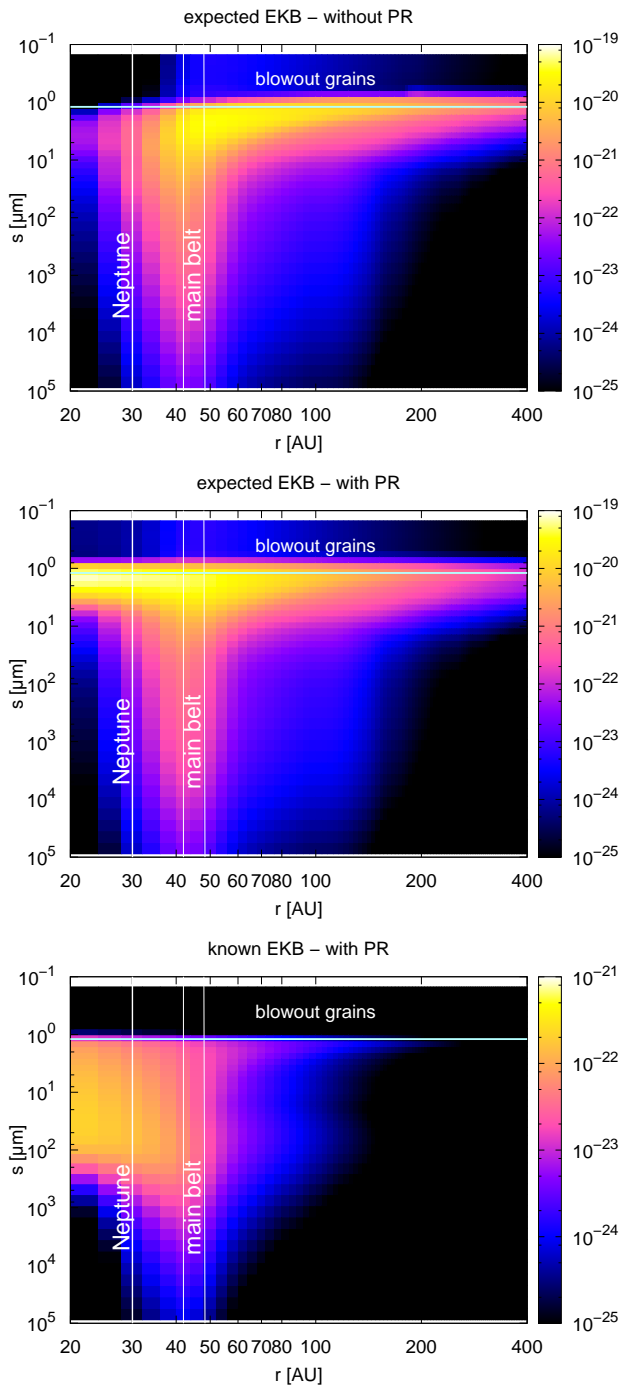}
  \end{center}
  \caption{
     The distribution of the cross section density of the EKB dust as a function
     of distance and grain sizes.
     The panels are as in Fig.~\ref{fig:size_different_distance}.
   }
  \label{fig:size_distance}
\end{figure}

Another view of the EKB dust can be achieved by plotting
its combined radial and size distribution (Fig.~\ref{fig:size_distance}).
Besides presenting the same salient features as those discussed in Sect.~3.2
and 3.3, it emphasizes that radial and size distribution of dust in a debris
disk are intrinsically coupled and cannot be treated
independently of each other.

\subsection{Optical depth}

\begin{figure}
  \begin{center}
  \includegraphics[width=0.48\textwidth]{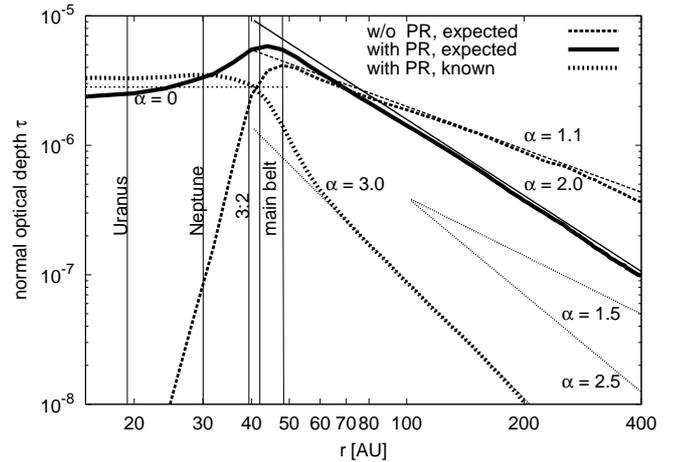}\\
  \end{center}
  \caption{Normal optical depth of the Kuiper belt with and without P-R drag.
  Uranus and Neptune are shown for orientation, but were not included in the simulations.
  The optical depth for the known EKB is amplified by a factor of $100$.
  }
  \label{fig:tau_compare}
\end{figure}

The  radial profile of the normal geometrical 
optical depth is shown in Fig.~\ref{fig:tau_compare}.
In the case of a disk dominated by the P-R effect, \citet{Strubbe-Chiang-2006} 
calculated analytically the exponent of the 
optical depth profile $\tau \propto r^{-\alpha}$ to be $\alpha = 2.5$ and 
$\alpha=0$ in the outer and inner regions, respectively.
Without P-R effect, i.e. for a collision-dominated disk,
no dust is present interior to the parent bodies, and the outer slope
should be close to $\alpha=1.5$.

These slopes are in qualitative agreement with our simulations
(Fig.~\ref{fig:tau_compare}).
Taking the known EKB objects as dust sources and including the P-R effect,
we find a nearly constant optical depth in the inner region and
a slope of $\alpha \approx 3.0$ in the outer disk, close enough
to predictions for a transport-dominated disk.
For the actual EKB dust disk and with the P-R effect taken into account,
the result is intermediate between what is expected for transport-dominated
and collision-dominated disks. This is seen from the inner profile
which is gently decreasing inward \citep[cf. Fig.~1 of][]{Wyatt-2005},
and from the outer slope of $\alpha \approx 2.0$.
Finally, for the actual EKB dust disk, but with the P-R effect switched off,
the profile is the one expected for collision-dominated disks.
The optical depth drops sharply inward from the main belt,
whereas in the outer region the slope is $\alpha \approx 1.1$.
That it is somewhat flatter than the analytic value $\alpha = 1.5$,
traces back to a rather broad radial distribution of scattered TNOs
that make a considerable constribution to the overall dust profile.

For the debiased EKB and with P-R effect included, 
the normal optical depth
peaks at $\approx 40\AU$ at a level
of $\approx 6 \times 10^{-6}$.
Besides the normal optical depth shown in Fig.~\ref{fig:tau_compare},
we have calculated the in-plane optical depth.
Our result, $\tau \leq 2\times 10^{-5}$ outside $30\AU$,
is at least by a factor of $4$ higher than the estimate by
\cite{Stern-1996} who found $\tau = 3\times 10^{-7}\dots 5\times 10^{-6}$.

We finally note that the dust production rate for the expected EKB in collisional 
equilibrium was calculated to $\lesssim 1.7\times 10^8\gpers$ and is three times 
higher than the predicted rate of $5\times 10^7\gpers$ from \cite{Landgraf-et-al-2002}
on the base of in situ measurements of Pioneer 10 and 11.

\section{Spectral energy distribution}

\subsection{Parameters and materials}

\begin{figure}[htb!]
  \begin{center}
  \includegraphics[width=0.48\textwidth]{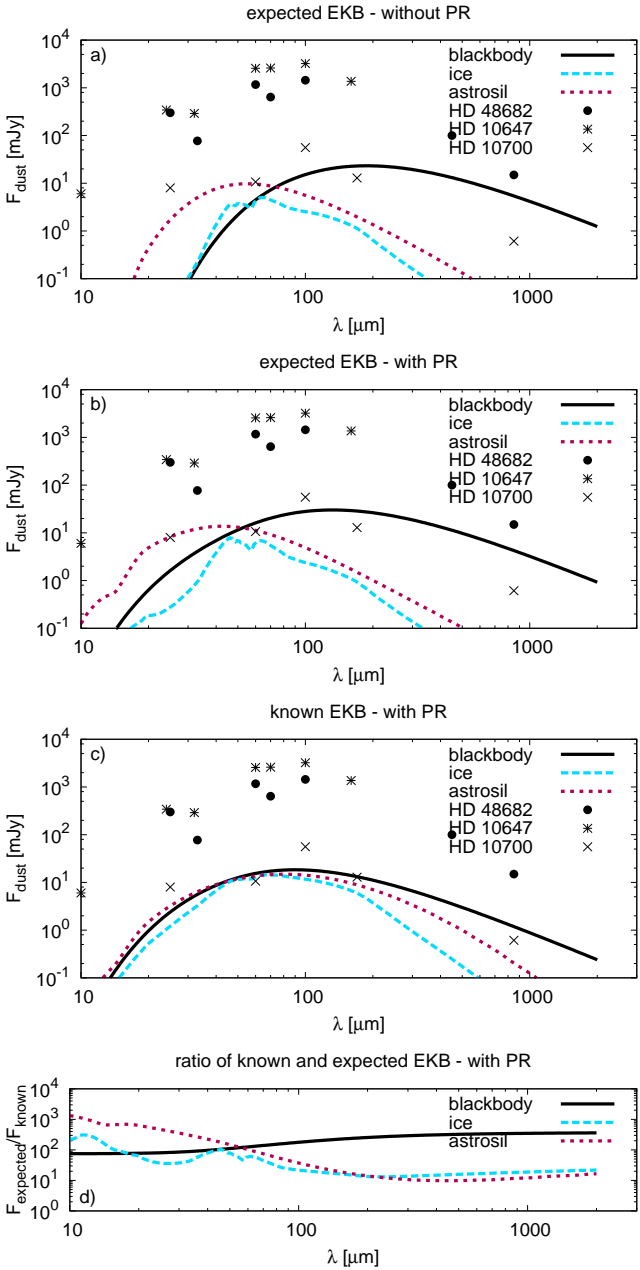}\\
  \end{center}
  \caption{Infrared excess of the Kuiper belt dust.
  (a): the debiased EKB, without P-R effect;
  (b): the debiased EKB, with P-R effect;
  (c): known EKB objects only, with P-R effect (amplified by a factor of $100$);
  (d): the ratio of (b) and (c).
  Three curves in each panel (a) to (c) are based on the same {\it ACE} runs, but the thermal emission
  was calculated
  for three different materials: blackbody, astrosilicate, and ``dirty ice''.
  For comparison, observed SEDs of three other old Sun-like stars are shown with symbols.
  Photometric data for these stars are adapted from 
  \citet{Greaves-et-al-2004,Sheret-et-al-2004,Beichman-et-al-2006,Chen-et-al-2006,Moor-et-al-2006,Lawler-et-al-2009,Tanner-et-al-2009}.
  }
  \label{fig:SED}
\end{figure}

The equilibrium temperatures of dust and their thermal emission were calculated
in a standard way as described, for instance, in \citet{Krivov-et-al-2008}.
In these calculations we computed the solar photospheric spectrum with the NextGen
grid of models \citep{Hauschildt-et-al-1999}, assuming
a G2V star of solar metallicity.
To get a rough idea of how the thermal flux is affected by (unknown) chemical composition
of the dust, we tested three different cases: ideal material (blackbody absorption and emission),
astrosilicate from \cite{Laor-Draine-1993}, and  ``dirty ice''.
The latter is defined as a mixture consisting of
a water ice matrix (with constants from \citet{Warren-1984}) contaminated with
$10\%$ volume fraction of astronomical silicate \citep{Laor-Draine-1993}.
This is similar to what, for instance, \citet{Yamamoto-Mukai-1998c} adopted
in their calculation of thermal emission of the EKB dust.
The refractive indices of the adopted ice-silicate mixture were calculated by means of 
the Maxwell-Garnett theory.
This also explains our using $10\%$ silicate fraction:
$10\%$ is an upper limit for which the effective medium theory still provides accurate 
results \citep{Kolokolova-Gustafson-2001}. 
With a standard Mie algorithm, we then computed
the absorption efficiency $Q_{abs}$ as a function of size and wavelength.

\subsection{Calculated flux}

The resulting SEDs of the simulated EKB dust disk, as it would be seen
from a $10\pc$ distance, are presented in Fig.~\ref{fig:SED}.
Three different curves correspond to three materials described above.
Pure astrosilicate and the ice-silicate
mixture produce SEDs of similar shape and height, peaking at $50$--$70\mum$
with a maximum flux at a level of several $\mJy$. 
On the ``dirty ice'' curve, a typical water ice feature at $\sim 60\mum$ is seen.
This feature may have been detected in the disk of a young debris disk star
HD~181327 \citep{Chen-et-al-2008}. The fact that the feature is located near
the SED maximum may help finding water ice in other debris disks by future
observations.
A comparison with the blackbody curve readily shows that it departs from the two others
significantly. The maximum moves to $\lambda \sim 200\mum$ and the (sub)-mm flux
becomes by two orders of magnitude higher than in the two other cases. This
result confirms earlier conclusion \citep[see, e.g.][]{Yamamoto-Mukai-1998c,Krivov-et-al-2008}
that the blackbody assumption is probably too crude and should not be used
in modeling the thermal emission of debris disks.

It is important to compare the calculated flux of the EKB dust with 
measurements of other known debris disks. For this purpose, we 
looked at  photometric data for other Sun-like stars.
We have chosen three stars which possess well-known, bright excesses
(measured at many wavelengths from mid-IR to sub-mm), which have spectral classes
not too far from solar G2, and which, like the Sun, are rather old
($\ga 1\Gyr$). These are
HD 48682 ($\psi^5$~Aur, G0V, at 17 pc distance, estimated age $0.6$--$9\Gyr$),
HD 10647 (q$^1$~Eri, F8V, 17 pc, $\sim 2\Gyr$), 
and HD 10700 ($\tau$~Cet, G8V, 4 pc, $7\Gyr$).
Their SEDs, normalized to $r = 10\pc$,
are overplottted in Fig.~\ref{fig:SED}.
A comparison shows that their SEDs are similar in shape to the simulated SED
of the EKB dust disk (for one of the two realistic materials, not blackbody).
The maximum of the SEDs of the selected debris disk stars
lies at $\sim 100\mum$, i.e. at slightly longer wavelenghts than the maximum
of the EKB dust flux. This (moderate) difference in the peak
wavelengths of different SEDs may be caused by a choice of chemical composition
of the disks, by different extentions of EKB and other debris disks, or both.

The major difference between the EKB dust disk and
extrasolar debris disks is, of course, the absolute level of the thermal emission
fluxes. For dust maintained by the debiased EKB (panels a and b), the fluxes
from far-IR to sub-mm are about one to two orders of magnitude
lower that those of the
reference stars. Obviously, this traces back to a much lower total mass of the EKB
compared to that of the extrasolar Kuiper belts. The reasons for this difference
are currently a matter of debate.
For instance, \citet{Booth-et-al-2009} argue that typical Kuiper belts around stars with
observed debris disks may not have undergone 
a major depletion phase due to a ``Late Heavy Bombardment''.
Currently, it is not known whether low-mass disks at the EKB level exist around other
stars at all and, if they do, how common they are.
This question can only be answered with future, more sensitive
observations (for instance, with Herschel Space Observatory).

It is interesting to compare fluxes from the expected EKB dust disk (Fig.~\ref{fig:SED}b)
with those from the dust disk that would stem from the known TNOs, without
debiasing. The corresponding SEDs, depicted in Fig.~\ref{fig:SED}c, are completely
different. Apart from an expected reduction of the fluxes by almost two orders of 
magnitude, two other effects are seen. One is that the astrosilicate and ``dirty ice'' SEDs
come much closer to each other and to the blackbody curve. As a result, their maxima
shift to longer wavelengths of $\sim 100\mum$. Another effect is a disapperance of the 
water ice feature. Both effects are easily explained by the major differences between
the size distribution of dust of the known and expected EKBs
(cf. middle and bottom panels in Fig.~\ref{fig:size}).
In the dust disk of the known EKB, the cross section, and thus the thermal emission, are 
dominated by grains $\sim 100\mum$ in size.
Such big grains behave as black bodies and do not produce
any distinctive spectral features.
We also present the ratio of fluxes  from the dust disks of the expected and known
EKBs (Fig.~\ref{fig:SED}d). The aforementioned reduction of the fluxes is seen here even
better. The reduction factor, $\sim 10$ at wavelengths longward of $\sim 100\mum$,
increases further at shorter wavelengths for ice-silicate and especially astrosilicate grains.
Obviously, this is also a consequence of the change in the size distribution
discussed above.

For three selected wavelengths of $70\mum$, $100\mum$, and $160\mum$, the fluxes
calculated for the ``known'' and ``expected'' EKBs and for all three materials
are also tabulated in Table~\ref{tab:flux}.
Solar photosphereic fluxes are given for comparison.
These values are used below to assess the detectability of debris disks with Herschel.

\subsection{Detectability of ``Kuiper belts''}

\begin{table}
\centering
\begin{tabular}{|c|c|c|c|c|c|c|c|}
\hline
 			& \multicolumn{2}{c|}{astrosil} & \multicolumn{2}{c|}{ice} 	& \multicolumn{2}{c|}{blackbody} & Sun\\
\hline
$\lambda$ [\textmu{}m]	& known & expected		& known & expected		& known & expected & \\
\hline
$70$			& $0.11$ & $7.9$		& $0.11$ & $4.8$			& $0.12$ & $17.3$ & $40$\\
$100$			& $0.11$ & $4.1$		& $0.10$ & $2.1$			& $0.14$ & $24.7$ & $22$\\
$160$			& $0.08$ & $1.5$		& $0.06$ & $0.9$			& $0.11$ & $25.5$ & $7.8$\\
\hline
\end{tabular}
\caption{Calculated flux of the known and expected Kuiper belt dust (including the P-R-effect) and photospheric flux of the Sun
 at different wavelengths from a $10 \pc$ distance. The fluxes are given in mJy.
 }
\label{tab:flux}
\end{table}

We would like to estimate the level, down to which Herschel can detect
faint debris disks, and then to compare that level with our EKB dust disk models.

The sensitivity limit of the PACS instrument of the Herschel telescope
at $100\mum$ in the scan-map mode for one hour exposure time
is $2.4\mJy$ at a $5\sigma$ uncertainty
level\footnote{\textit 
{http://herschel.esac.esa.int/Docs/PACS/html/ch03s05.html\#sec-photo-sensitivity}},
or $0.48\mJy$ at $1\sigma$.
The background noise, of course, is highly variable from one star to another.
For $133$ target stars of the Open Time Key Program DUNES (DUst around NEarby 
Stars), its average value at $100\mum$ is $0.53\mJy$ at $1\sigma$.
Combining the instrument and the background noise leads to a limiting flux
as low as $5\sqrt{0.48^2+0.53^2} = 3.6\mJy$ for one hour exposure time at a $5\sigma$ uncertainty level.
Assuming that the $100\mum$ flux is proportional to the total mass of a debris disk,
we can conclude that an extrasolar EKB analog with a mass $M > 1.7 M_\text{EKB}$
will be detectable from a distance of $10\pc$, assuming a ``dirty ice'' as a dust composition.
For silicate dust, the detectability limit would go down to $0.9 M_\text{EKB}$.
Additional uncertanties in the stellar photospheric flux
($\approx 2\%$ of $22\mJy$, or $0.4\mJy$) increase these
values slightly to $\approx 1.9 M_\text{EKB}$ and $\approx 1.0 M_\text{EKB}$, respectively.

\section{Summary and conclusions}

In this paper, we attempted to construct a model of the solar system
debris disk~--- Kuiper belt objects and their collisional debris.
This was done in two major steps.
First, we developed a new algorithm to remove the two largest selection effects from
TNO observations, the inclination and the distance biases.
Applying it to the database of 1260 known TNOs, we derived new estimates for
the global parameters of the ``true'' Kuiper belt: its mass, orbital and
size distribution.
Second, treating the debiased populations
of EKB objects as dust parent bodies, we employed our collisional code
to simulate their dust disk.
This has resulted in estimates of 
the size and radial distribution of dust in the transneptunian region,
as well the optical depth of the Kuiper belt dust disk.
We have also calculated the thermal emission of the Kuiper belt dust
and compared the expected infrared fluxes with those of known debris disks
around other Sun-like stars.

\bigskip
We draw the following conclusions:
\begin{enumerate}
\item
The total mass of the EKB, including classical, resonant, and scattered objects,
amounts to $0.12 M_\oplus$.
Without scattered objects, the mass reduces to 
$0.05 M_\oplus$.
These values are by about one order of magnitude higher than the
mass of the known TNOs.
\item
The dust disks that would be produced by the known EKB
and by the expected (i.e., debiased) one are shown
to have distinctly different properties,
as detailed below.
Much of the difference in the size and radial distribution of dust
of the two EKBs traces back to the increased total mass of TNOs
after the debiasing.
\item
The slope of the differential size distribution in the EKB
in the size range $50\km < s < 170\km$ (absolute magnitudes $8.9>H>6)$
is found to be $q = 4.3\pm 0.2$ for the classical objects,
$5.1\pm 0.1$ for resonant ones,
and $2.8\pm 0.1$ for scattered TNOs.
This results in $4.4\pm 0.2$ for classical and all resonant objects together.
For the entire EKB (classical, resonant, and scattered TNOs),
we find $3.6\pm 0.1$.
\item
If the debiased populations of TNOs are taken as dust sources,
the dustiness of the disk is high enough to make collisional timescales
comparable to, or even shorter than, the radial transport timescales.
Thus the predicted Kuiper belt dust disk rather falls to the category of
collision-dominated debris disks, to which all debris disks detected
so far around other stars belong, albeit it is already close to the ``boundary''
between collision-dominated and transport-dominated disks.
The simulated size distribution in the EKB dust disk shows a sharp cutoff
at the radiation pressure blowout radius of $\sim 1\mum$. The cross section
dominating radius is several times larger.
\item
For comparison, we have also considered the dust disk that would be produced solely by 
known TNOs as parent bodies.
That disk would fall into the transport-dominated regime,
where the Poynting-Robertson drift timescales are shorter than the collisional
timescales (at dust sizes).
The size distribution in such a disk is dramatically different from
the one expected in collision-dominated disks. While a cutoff at $1 \mum$
remains, the distribution between $1\mum$ and several hundreds of $\mum$
is nearly flat, and the cross section is now dominated by much larger
grains, $\sim 100\mum$ in radius.
\item
The radial distribution of the TNOs themselves
peaks in the region of the main belt ($40$--$50\AU$),
where about 90\% of the cross section comes from the classical EKB
objects.
Outside $\sim 60\AU$, the cross section is solely due to scattered objects.
The distribution of the latter is quite extended radially, being nearly flat
over a wide distance range from $\approx 35\AU$ to more than $100\AU$.
\item
The radial distribution of dust grains with radii $\ga 10\mum$ is similar
to the distribution of the parent bodies described above.
At smaller sizes, the radial profile gets progressively broader with decreasing
radius, which is a classical effect of radiation pressure.
\item
The maximum normal geometrical optical depth is reached at the inner edge
of the classical belt, $\approx 40\AU$, and is estimated to be
$\approx 6 \times 10^{-6}$. Outside that distance, it falls off as
$r^{-2}$. This slope is roughly intermediate between the slope
predicted analytically for collision-dominated ($r^{-1.5}$)
and transport-dominated ($r^{-2.5}$) disks.
An upper limit of the in-plane optical depth is set to $\approx 2\times 10^{-5}$ outside $30\AU$.
\item
For comparison, the normal optical depth of the
dust disk produced only by known TNOs would fall off
outside the classical EKB as $\tau \propto r^{-3}$, and the
in-plane optical depth would be by two orders of magnitude lower.
\item
The estimated thermal emission flux from the EKB dust disk that would be
observed from a $10\pc$ distance strongly depends on the assumed dust composition.
For two ``realistic'' materials probed, astrosilicate and ice-silicate mixture,
the SED would peak at $40$--$60\mum$. The maximum value is at a level of
several $\mJy$ (at the solar photospheric flux at the same wavelengths
is several tens of $\mJy$), 
which is about two orders of magnitude smaller than for 
the brightest known debris disks around other solar-type stars,
normalized to the same distance.
For a non-debiased EKB, the fluxes would be another two orders of magnitude
lower.
\item
With the Herschel Space Observatory, it may be possible to detect faint 
debris disks around other stars, which are nearly as faint as the EKB dust disk.
For observations with the PACS instrument at $100\mum$ in a scan-map mode with one hour exposure, 
assuming an average background noise, the minimum mass of a Kuiper belt analog detectable
at $5\sigma$ from $10\pc$ is estimated as $\approx 1\dots 2 M_\text{EKB}$.
\end{enumerate}

After the submission of this paper, we became aware of another
work \citep{Kuchner-Stark-2010} that addresses the same
problem of predicting the EKB dust disk properties and, like this paper,
employs a catalog of known KBOs to model the dust production.
Instead of collisional grinding simulations,
\citeauthor{Kuchner-Stark-2010} make use of N-body integrations
supplemented with a ``collisional grooming'' algorithm.
Despite the difference in the simulation methods, their
results for the radial distribution of the EKB dust are in 
a very good agreement with ours
(cf. our Fig.~\ref{fig:size_different_distance} and their Figs.~4--5).
The same holds for size distributions at lower optical depth levels
where they find the same turn-off in the size distribution  as we do
(cf. our Fig.~\ref{fig:size} and their Fig.~6).
As a result, the size of cross section-dominating grains
in the densest parts of disks becomes larger.
Since these larger grains are cooler, the
maximum of the SED should move to longer wavelengths with
decreasing $\tau$
(cf. our Fig.~\ref{fig:SED} and their Fig.~9).

\section{Discussion}

The main result of this paper is a set of new estimates of the parameters
of the ``true'' (debiased) Kuiper belt and its debris dust. As all models,
ours involves uncertainties. Here we discuss three major ones.

{\em 1. Parent bodies.}
Although we believe that our debiasing algorithm is more accurate that
the ones used before, it is possible that the results over- or underestimate,
for instance, the mass of the EKB by a factor of a few.
The main reason is an incomplete list of surveys used for debiasing.
In particular, observations that covered higher ecliptic latitudes,
omitted in Tab.~\ref{tab:observations},
could alter the results.
We have identified 90 objects in the MPC
(excluding Centaurs) that have been detected beyond $10^\circ$ ecliptic
latitude. Of these, 29 objects have been found beyond $20^\circ$.
To check how non-including the surveys that covered higher latitudes
may affect the results, we have made additional tests. We have
re-run our code several times, each time artificially changing $\epsilon$
in three randomly chosen surveys of Tab. \ref{tab:observations}
to $20^\circ$.
This changed the detection probability of single objects typically by
a few tenths of percent only. The change in the inferred total EKB mass
turned out to be larger, but still minor, 4\%.

Another difficulty is related to the scattered objects
with their large eccentricities and inclinations.
Their average detection 
probability is very low, the resulting ``mass amplification factor''
very high ($\approx 25$), so that the estimated masses of objects can
considerably differ from the ``true'' ones.

The same applies to the treatment of the smallest objects discovered so far,
a few kilometers is radius. If, for instance, the break
the size distribution at $s\approx 30\km$ inferred by
\citet{Bernstein-et-al-2004} and \citet{Fraser-2009}
were included, this would reduce the overall amount of dust.
This would make the EKB dust disk ``more transport-dominated''
with all the consequences disussed above.

Being aware of such caveats, throughout the paper
we always compare the known (not debiased) EKB with the expected (debiased) EKB.
This applies to the EKB itself (Sect.~2), as well as to its dust (Sect.~3--4).
We expect that
the results obtained here for the known and debiased EKB
would at least "bracket" the reality.

{\em 2. Material composition of solids.}
Realistic simulations would necessitate good knowledge of tens
of parameters, such as the bulk density, porosity, shape, tensile strength,
optical constants of solids in the EKB, and all this over an extremely
broad range of sizes: from hundreds of kilometers (large TNOs) down to a fraction of
a micrometer (tiniest dust grains). In this paper, we made
a large set of simplifying assumptions, ignoring, in particular, any dependence
of all these parameters on size, although such dependencies are expected.
We only probed several different materials when calculating the thermal
emission of dust in Sect.~4, and saw how sensitive the results are to
the adopted absorption and emission efficiencies of grains.
There are no reasons to think that variation of other parameters, for instance
mechanical properties of different-sized solids in the collisional modeling,
would alter the results to a lesser extent.

{\em 3. Modification of dust disk by Neptune.}
As mentioned above, effects of giant planets on the dust distribution,
of which those by Neptune are the largest,
were not included in our simulations.
These effects are diverse and can alter the results substantially.
Neptune is expected to capture some grains stemming from non-resonant
KBOs and drifting inward by P-R effect
into mean-motion resonances
\citep[e.g.][]{liou-zook-1999,MoroMartin-Malhotra-2002}.
Similarly, sufficiently large grains that derive from resonant TNOs 
must stay locked in such resonances from the very beginning 
for considerable time periods
\citep[e.g.][]{Wyatt-2006,Krivov-et-al-2007}.
This will cause an enhancement of dust density at resonance locations,
which was not modeled here.
Furthermore, the distribution of dust will exhibit azimuthal
clumps instead of being rotationally-symmetric, as was assumed in our
{\it ACE} simulations.
Next, many of the grains are in Neptune-crossing orbits, or reach such
orbits in the process of their inward drift. The probability of close encounters
with Neptune is not negligible, especially for larger grains that
drift inward more slowly. Encounters would result in partial truncation
of the dust disk at the planet's orbit, i.e. in a drop of dust density
at $30\AU$. Some of the grains will be ejected outward, which may affect
the collisional balance of the whole disk.

For these reasons, the simulation results presented here should be treated
as preliminary. We can hope that future, deeper,
TNO surveys would lead to a better knowledge of the global parameters
of the EKB acting as dust parent bodies.
Further modeling work, perhaps with other methods, should help quantify
the effects of Neptune on the dust distribution. Choosing the most adequate
mechanical and optical properties of KBOs and their dust is probably the
main difficulty. Eventually, many of the uncertainties could be minimized,
and the models can be verified,
if dust in the transneptunian region is detected in situ by dust detectors
aboard space missions, such as New Horizons
\citep{Horanyi-et-al-2008}.
On any account, we consider this study as a reasonable starting point
in developing more accurate models of the debris disk of our own solar system.

\begin{acknowledgements}
We thank Mark Wyatt for stimulating discussions of several aspects of this work
and useful suggestions, Sebastian M\"uller for assistance in thermal
emission calculations presented in Sect.~4,
and the anonymous reviewer for a useful referee report.
This research was supported by the
\emph{Deut\-sche For\-schungs\-ge\-mein\-schaft} (DFG), project number Kr~2164/8-1.
AK and TL are grateful to the Isaac Newton Institute for Mathematical Sciences
in Cambridge where part of this work was carried out in the framework
of the program ``Dynamics of Discs and Planets''.
\end{acknowledgements}

%------------------------------------------------------------------
% Bibliography
%------------------------------------------------------------------

\newcommand{\AAp}      {Astron. Astrophys.}
\newcommand{\AApSS}    {AApSS}
\newcommand{\AApT}     {Astron. Astrophys. Trans.}
\newcommand{\AdvSR}    {Adv. Space Res.}
\newcommand{\AJ}       {Astron. J.}
\newcommand{\AN}       {Astron. Nachr.}
\newcommand{\AO}       {App. Optics}
\newcommand{\ApJ}      {Astrophys. J.}
\newcommand{\ApJS}     {Astrophys. J. Suppl.}
\newcommand{\ApJL}     {Astrophys. J. Letters}
\newcommand{\ApSS}     {Astrophys. Space Sci.}
\newcommand{\ARAA}     {Ann. Rev. Astron. Astrophys.}
\newcommand{\ARevEPS}  {Ann. Rev. Earth Planet. Sci.}
\newcommand{\BAAS}     {BAAS}
\newcommand{\CelMech}  {Celest. Mech. Dynam. Astron.}
\newcommand{\EMP}      {Earth, Moon and Planets}
\newcommand{\EPS}      {Earth, Planets and Space}
\newcommand{\GRL}      {Geophys. Res. Lett.}
\newcommand{\JGR}      {J. Geophys. Res.}
\newcommand{\JQSRT}    {J. Quantitative Spectroscopy and Radiative Transfer}
\newcommand{\MNRAS}    {MNRAS}
\newcommand{\PASJ}     {PASJ}
\newcommand{\PASP}     {PASP}
\newcommand{\PSS}      {Planet. Space Sci.}
\newcommand{\SolPhys}  {Sol. Phys.}
\newcommand{\SolSysRes}{Sol. Sys. Res.}
\newcommand{\SSR}      {Space Sci. Rev.}

\end{document}